\documentclass [a4paper,12pt]{article}
\usepackage{amsmath,comment,url}
\usepackage{amsfonts,mathtools}
\usepackage{amssymb,float,bm}
\usepackage[pdftex]{graphicx,color} 
\usepackage{latexsym}
\usepackage[title]{appendix}
    \newenvironment{dedication}
        {\vspace{6ex}\begin{quotation}\begin{center}\begin{em}}
        {\par\end{em}\end{center}\end{quotation}}

\mathtoolsset{showonlyrefs}

\def\b{\begin{eqnarray}}
\def\e{\end{eqnarray}}

\def\ii{\int\limits_{\mathbb{R}}}

\def\u{\mathfrak{u}}
\def\eps{\varepsilon}
\def\L{\mathfrak{L}}
\graphicspath{{Figures/}}
\usepackage{authblk}

\newtheorem{remark}{Remark}[section]

\newcommand{\R}
{\mathbb{R}}

\newcommand{\Beq}{\begin{equation}}
\newcommand{\Eeq}{\end{equation}}

\newcommand{\BS}{\begin{subequations}}
\newcommand{\ES}{\end{subequations}}
\newcommand{\Beqn}{\begin{equation*}}
\newcommand{\Eeqn}{\end{equation*}}
\newcommand{\Beqa}{\begin{eqnarray}}
\newcommand{\Eeqa}{\end{eqnarray}}
\newcommand{\Beqan}{\begin{eqnarray*}}
\newcommand{\Eeqan}{\end{eqnarray*}}

\title{Higher-order integrable models for oceanic internal  wave-current interactions}
\author[1]{David Henry \thanks{d.henry@ucc.ie} }
\author[2]{Rossen I. Ivanov \thanks{rossen.ivanov@tudublin.ie}}
\author[1]{Zisis N. Sakellaris \thanks{zsakellaris@ucc.ie}}

\affil[1]{\small{School of Mathematical Sciences, University College Cork, Cork T12 XF62, Ireland}}
\affil[2]{\small{School of Mathematics and Statistics, Technological University Dublin, Grangegorman Lower, Dublin D07 ADY7, Ireland}}

\date{}

\begin{document}
\maketitle

\vspace{-4em}

\begin{dedication}
{Dedicated to the memory of Prof. David J. Kaup (1939-2022)}
\end{dedication}

\begin{abstract}
\noindent In this paper we  derive a higher-order KdV equation (HKdV) as a model to describe the unidirectional propagation of waves on an internal interface separating two fluid layers of varying densities. Our model  incorporates underlying currents by permitting a sheared current in both fluid layers, and also accommodates the effect of the Earth's rotation by including Coriolis forces (restricted to the Equatorial region).    The resulting governing equations describing the water wave problem in two fluid layers under a `flat surface' assumption are expressed in a general form as a system of two coupled equations through Dirichlet-Neumann (DN) operators.  The DN operators also facilitate a convenient Hamiltonian formulation of the problem. We then derive the HKdV equation  from this Hamiltonian formulation, in the long-wave, and small-amplitude, asymptotic regimes. 
Finally, it is demonstrated that there is an explicit transformation connecting the HKdV we derive with the following integrable equations of similar type: KdV5, Kaup-Kuperschmidt equation, Sawada-Kotera equation, Camassa-Holm and Degasperis-Procesi equations.
\end{abstract}

\begin{section}{Introduction}
In this paper we derive  a Higher-order KdV (HKdV) equation from a physical model which describes the unidirectional propagation of waves on an internal interface separating two fluid layers of varying densities.  Our model accommodates the effects of the Earth's rotation by way of Coriolis forces,  and we also incorporate underlying currents by permitting a sheared current in both fluid layers. 

Geophysical fluid dynamics (GFD) concerns the motion of large scale fluid flows in the ocean, or atmosphere, and the two primary distinguishing features of GFD compared to traditional fluid mechanics are the inclusion of the effects of Earth's rotation, and of stratification \cite{CRB}.  
Stratification is an important effect when fluids of different densities interact, as is the case for many naturally occurring scenarios involving, for example: cool and warm water (or air); fresh and brackish water.  In equilibrium under gravity, fluid layers are stably stratified by either forming into a series of distinct horizontal layers consisting of fluid of the same density, or by a continuous stratification. In this paper we consider only the former scenario, in a model whereby two homogeneous fluid layers  of different  densities are separated by an interface known as a pycnocline. When density differences are due to marked variations in temperature, the pycnocline coincides with the thermocline. A simplified schematic for these waves is presented in Figure \ref{fig:my_label}. Internal waves are generated when fluid disturbances disrupt  the equilibrium, with energy then propagating along the pycnocline. 

\begin{figure}[h!]
    \centering
    \includegraphics[width=.6\textwidth]{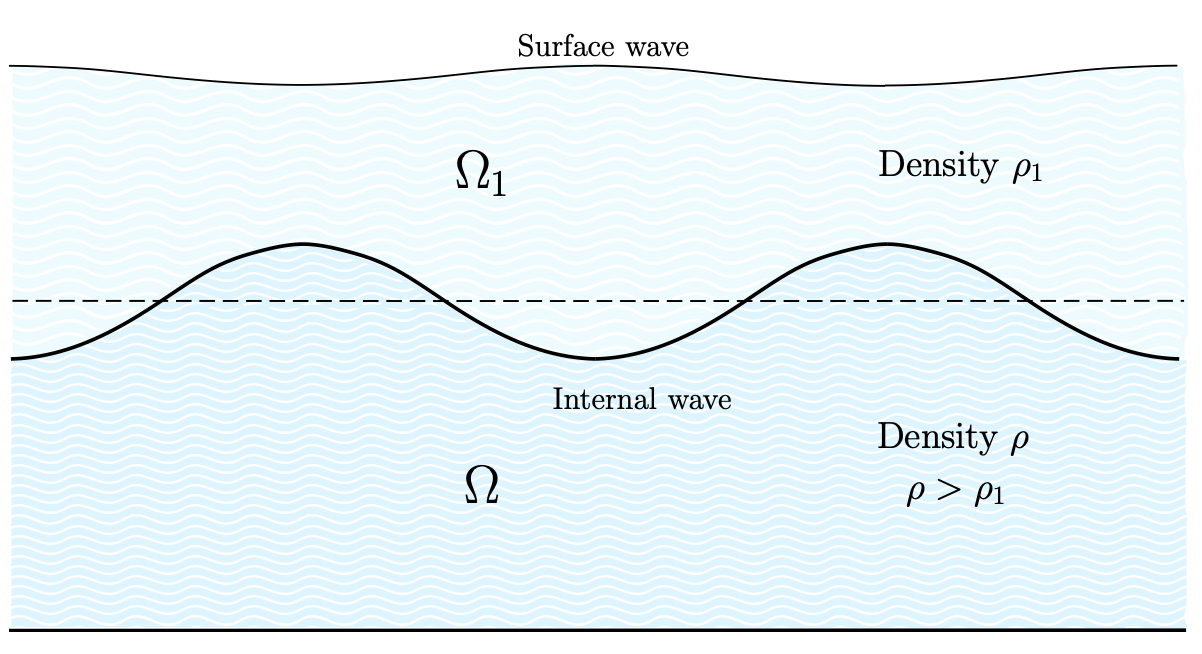}
    \caption{Schematic of a two-layer fluid model of an interfacial internal wave.}
    \label{fig:my_label}
\end{figure}

A significant complicating factor in the model we consider results from our inclusion of underlying currents in the form of a sheared current in both fluid layers. Such underlying currents play an important role in various geophysical phenomena, however they have typically been  neglected in the literature when deriving integrable model equations  due to their  introducing non-zero vorticity into the flow kinematics, which in turn greatly complicates the mathematical formulation of the water wave problem. A germane example of an underlying current which interacts with internal waves propagating on the pynocline is the Equatorial Undercurrent (EUC), which is located  within approximately 200 -- 300 km (below 3$^{\circ}$ latitude) of the Equator. The EUC is symmetric about the Equator and extends nearly across the entire extent (more than 12000 km) of the Pacific Ocean basin \cite{Izumo}. The speed of the EUC is usually slightly over 1 m/s, which makes it one of the fastest permanent currents in the world.
The equatorial region in the Pacific is also characterised by a thin shallow layer of warm and less
dense water over a much deeper layer of cold denser water. The two layers are separated by a sharp
thermocline at a depth which depends on the location, but which typically lies at 100 -- 200 m beneath the surface \cite{Sim}. Both layers are to a large degree homogeneous and their sharp boundary (interface) is the thermocline/pycnocline. Aside from incorporating stratification and underlying currents, our model also accommodates the effect of the Earth's rotation by including Coriolis forces in the governing equations of motion, although we restrict our focus to the Equatorial region. Although classically neglected in the literature, we refer to \cite{IC,CIT,CI15,CI,ConIv,EHKL,HI,IV09,IMT,RJ}  for various  approaches which do incorporate shear currents, or Coriolis forces, in the derivation of integrable model equations.

With regard to modelling considerations, in the following the velocity field will be composed of wave-like, and current, components. It will be assumed that the underlying current is sheared in each fluid layer, that is, it  varies linearly with the depth within each layer (or, more properly, in fluid strips  excluding the region encompassing the internal wave interface). This assumption is equivalent to the vorticity of the flow in each of the layers being constant (thus automatically satisfying the vorticity equation) implying that the dynamics on the interface reduce to a single Bernoulli-type equation \cite{IC,CI,CIM,CuIv2}.  We also invoke a flat-surface (akin to a `rigid-lid') assumption, which is applicable in physical settings whereby the motion at the interface does not create a significant disturbance at the surface (cf. \cite{HV1,HV2} for a detailed mathematical examination of the various coupling effects which may otherwise occur between the internal and surface waves, and the resulting underlying flow kinematics).

The first main development of this paper is to recast the nonlinear governing equations for internal wave-current interactions as a closed system of coupled equations in terms of Dirichlet-Neumann (DN) operators for the upper and lower fluid layers, and for variables restricted to the interface. The governing equations are inherently nonlinear, but can be subjected to various approximations using the convenient setting of the Hamiltonian formulation of the problem \cite{Compelli1Wavemotion,IC}. We focus on the small-amplitude,  and long-wave, asymptotic approximations --- the so-called `KdV regime' \cite{KDV} --- noting that in an oceanographical context the long-wave regime may be considered the most important propagation regime since wave energy tends to be concentrated in the lower frequencies. KdV is the best known model equation describing the competition between nonlinear and dispersion effects in nonlinear media, see for example \cite{Johnson,ZMNP} or, in the context of internal waves \cite{IC}.  
 In this paper we include asymptotic expansions with terms of higher order than the standard  KdV approximation, resulting in the derivation of a Higher-order KdV  (HKdV) equation for the propagation of internal waves with currents. We relate the derived HKdV equation to all known integrable equations with the same type of nonlinear and dispersive terms using a so-called `Near Identity' transformation, namely the: KdV-5 equation (with $5^{th}$ derivative) from the KdV hierarchy, the Kaup-Kuperschmidt equation, the Sawada-Kotera equation, and also the integrable equations in non-evolutionary form, the Camassa-Holm and Degasperis-Procesi equations. 

 The paper is organised as follows. In Section \ref{sec2} we introduce the governing equations and perform their reduction to a coupled system of two equations on the interface, namely a Bernoulli-type equation, and the kinematic boundary condition on the interface. In Section \ref{sec:DNO} we introduce the DN operators for the fluid domains and we express the equations for the interface in terms of the DN operators. In Section \ref{sec:Ham} the equations are reformulated in Hamiltonian form. Then the scale parameters are introduced in Section \ref{sec:asympt} and the asymptotic expansion of the Hamiltonian in terms of the small-scale parameters is provided. The HKdV equation is derived from the Hamiltonian formulation in Section \ref{sec:HKdV}. The relation of the HKdV to the integrable models via a Near-Identity transformation is presented in Section \ref{sec:Int}.

\section{Governing equations and preliminaries}\label{sec2}
 The governing equation of motion for inviscid geophysical fluid dynamics  is given by the Euler equation
\begin{equation} \label{E}
{\bf V}_t+ ({\bf V} \cdot \nabla){\bf V}+ 2 \bm{ \omega} \times {\bf V}=-\frac{1}{\rho}\nabla P -{\mathbf{g}}, \end{equation} where  ${\bf V}$ represents the velocity field and $\bm{ \omega}$ is the Earth's angular velocity --- with $\bm{ \omega} \times {\bf V}$ the resulting Coriolis force, $\mathbf{g}=(0,0,g)$ represents acceleration due to gravity, and $P$ is the pressure in the fluid. 
Taking a Cartesian coordinate system whose origin is located on the Earth's surface at latitude $\theta$ --- with the $x-$axis oriented to the East, the $y-$axis pointing due North, and the $z$ axis vertical to the Earth's surface --- then  $\bm{ \omega} =\omega(0, \cos \theta, \sin \theta)$, where $\omega=7.3\times 10^{-5}$ rad/s, and  $2  \bm{ \omega} \times {\bf V}=(\hat f  V_z  - f V_y, f V_x,-\hat f V_x)$ where ${\bf V}=(V_x,V_y,V_z)$ denotes the Cartesian components of the velocity vector, $f=2\omega \sin \theta$ is the Coriolis parameter, and $\hat f=2 \omega \cos \theta$. In the equatorial region the Coriolis forces are directed towards the equator in both hemispheres, with the equator acting effectively  as a natural waveguide. Coriolis forces take a particularly simple form at the equator, $2  \bm{ \omega} \times {\bf V}=(\hat f  V_z, 0,-\hat f V_x)$,  with no forces acting in the $y$ (North-South) direction. Accordingly, in the equatorial region we may reasonably restrict ourselves to two-dimensional fluid motion moving zonally along the equator: we implement this restriction in what follows.

\begin{figure}[htp]
    \centering
    \includegraphics[width=.7\textwidth]{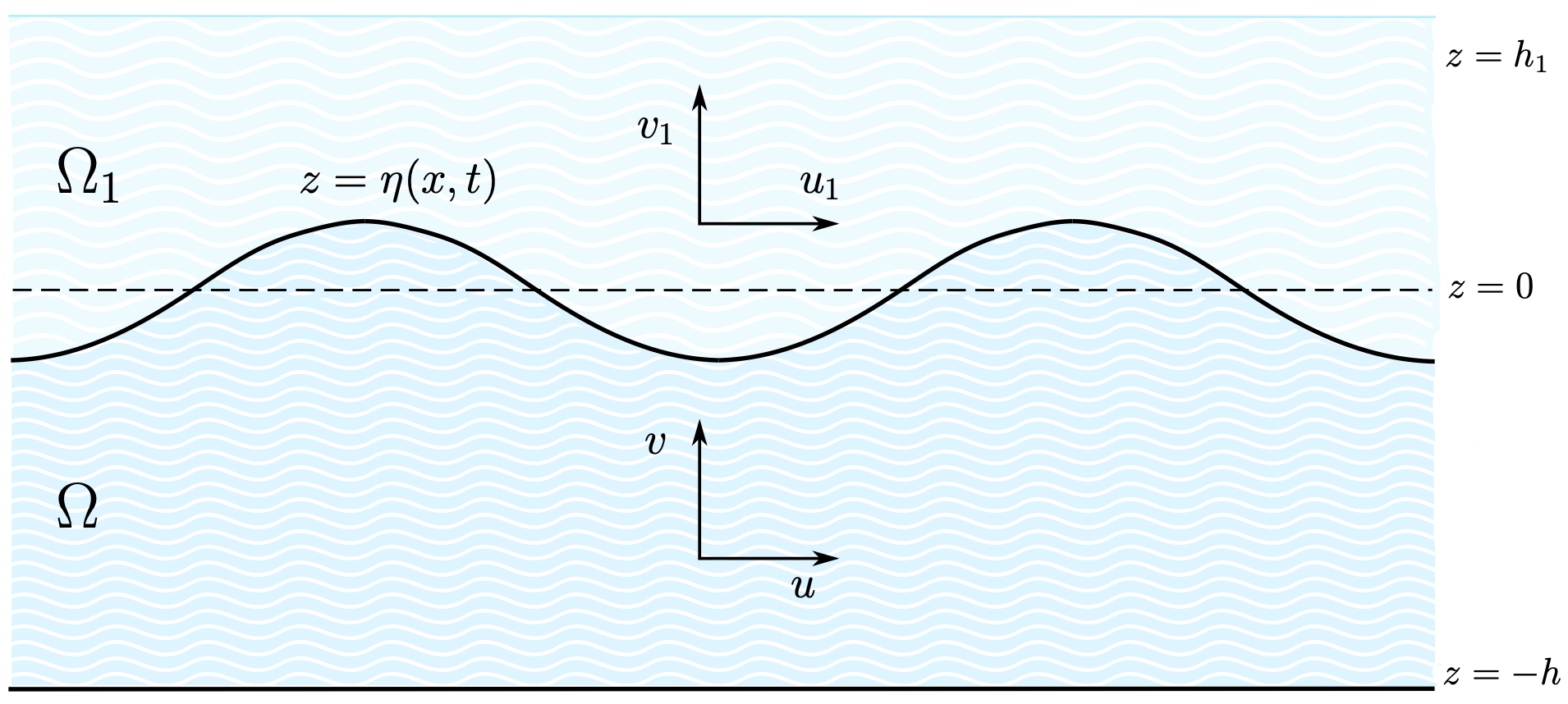}
    \caption{The fluid domains of the two layer model with a flat bottom in the flat surface approximation.}
    \label{fig}
\end{figure}

The fluid domain in the $(x,z)$ plane is shown in Figure \ref{fig}, $x$ is the horizontal coordinate running from West to  East, while $z$ is the vertical coordinate. It consists of two parts: the lower fluid domain,  $\Omega:=\left\{ (x,z) \in \R^2: x \in \R, -h \leq z \leq \eta(x,t) \right\}$, and the upper fluid domain, which is defined as $\Omega_1:=\left\{ (x,z) \in \R^2: x \in \R, \eta(x,t) \leq z \leq h_1 \right\}$, with $t$ being the time variable.  In this setting, the flat bed is located at $z=-h$ and the flat surface at $z=h_1$.  We use $\eta $ to describe the deviation of the free interface from its equilibrium state $\eta \equiv 0.$
 In physical terms, the interface $z=\eta(x,t)$ describes the location of the pycnocline (or thermocline), separating the two fluid layers, so by definition the average deviation is zero, that is  
\[
\int_{\R} \eta(x,t) dx = 0,
\] 
for all $t$. The fluids are incompressible and with constant densities $\rho$ and $\rho_1$ of the lower and upper media respectively. For immiscible fluids, the stability of the motionless equilibrium state with $\eta \equiv 0$  is ensured by the condition
$\rho>\rho_1$.
The two-dimensional velocity fields in the upper, and lower, layer will be denoted by ${\bf{V}}_1=(u_1(x,z,t),v_1(x,z,t))$, and ${\bf{V}}=(u(x,z,t),v(x,z,t))$, respectively. In order to model a depth-dependent shear current in both layers, we decompose  $u={\varphi}_{x} +\gamma z+\kappa$, and $u_1  =  {\varphi}_{1,x} +\gamma_1 z+ \kappa_1 $, where $\gamma=u_z-v_x$ and $\gamma_1=u_{1,z}-v_{1,x}$ are the constant non-zero vorticities in the two media; $\kappa $ and $\kappa_1 $ are the constant mean-horizontal velocity components of the flows; and ${\varphi}$ and ${\varphi}_1$ are velocity potentials which prescribe the wave-like motion. The velocity field can then be expressed as
\begin{equation}
\label{def__stream_phi}
        \begin{array}{lcl}
  \left.        \begin{array}{lcl}
        u  =   {\varphi}_{x} +\gamma z+\kappa=\psi_z(x,z,t)
        \\
        v=  { {\varphi}}_{z}=-\psi_x
        \end{array}\right\}  \quad \mbox{in $\Omega$}
\\
  \left.        \begin{array}{lcl}
        u_1  =  {\varphi}_{1,x} +\gamma_1 z+ \kappa_1=\psi_{1,z}
        \\
        v_1=  { {\varphi}}_{1,z}=-\psi_{1,x}
        \end{array}\right\} \quad \mbox{in $\Omega_1$}
        \end{array}
\end{equation}
where $\psi$ and $\psi_1$ are stream functions for the respective fluid layers. For simplicity we assume that $\kappa=\kappa_1$.
The decomposition \eqref{def__stream_phi} separates the fluid velocity into  ``wave-motion'' components, determined by the potentials ${\varphi}$ and ${\varphi}_1$, and underlying sheared currents $\gamma z +\kappa $ and $\gamma_1 z + \kappa_1 $ in the corresponding domains. 
An immediate consequence of \eqref{def__stream_phi} is that the fluid is incompressible in both layers, that is,  $\nabla\cdot {\bf{V}}\equiv \nabla\cdot {\bf{V}}_1\equiv 0$, if and only if ${\varphi}$ and ${\varphi}_1$ are both harmonic functions:
\begin{equation}
    \Delta \varphi= \varphi_{xx}+\varphi_{zz}=0, \quad \Delta \varphi_1= \varphi_{1,xx}+\varphi_{1,zz}=0.
\end{equation}
\begin{remark}
    By definition of \eqref{def__stream_phi}, the Cauchy-Riemann equations 
    \begin{equation}
        \varphi_x = \Psi_z, \quad \varphi_z=-\Psi_x
    \end{equation}
     are satisfied in the domain $\Omega$, where $\Psi:= \psi - \frac{1}{2}\gamma z^2 - \kappa z $. Similar Cauchy-Riemann equations are satisfied in $\Omega_1$. As a consequence, the velocity field in both domains can be recovered from the boundary values of the velocity potential.   
\end{remark}
\noindent The boundary conditions on the surface and at the bottom are
\begin{equation}
    v_1(x,h_1)=\varphi_{1,z}(x, h_1)=0 \, \, \text{and} \,\,  v(z,-h)=\varphi_z(x,- h)=0.
\end{equation}
In what follows, the subscript $c$ will be used for any expression evaluated on the common interface $z=\eta(x,t).$ The kinematic boundary condition on the interface is 
  \[
     \eta_t  =-\eta_x (u)_c + \left( v \right)_{c}=  - \eta_x (u_1)_c + \left( v_1 \right)_{c},
     \]
  which can be restated as 
\begin{equation}\label{KBC}
     \eta_t  =-\eta_x \left[   \left( \varphi_{x} \right)_{c} + \gamma  \eta + \kappa \right]  + \left( \varphi_{z} \right)_{c}=  - \eta_x \left [  \left( \varphi_{1,x} \right)_{c}+ \gamma_1  \eta + \kappa  \right ] + \left( \varphi_{ 1, z} \right)_{c}.
\end{equation}
Due to the localised nature of the wave forms under consideration, we assume that the functions $\eta(x, t),$ ${\varphi}(x, z, t)$ and ${\varphi}_1(x, z, t)$ belong to the Schwartz class $\mathcal{S}(\mathbb{R})$ with respect to the $x$ variable (for any $z$ and $t$), which implies that for large  values of $x$ the internal wave profile diminishes \begin{equation}\label{sch}
\lim_{|x|\rightarrow \infty}\eta(x,t)=0, \quad \lim_{|x|\rightarrow \infty}{ {\varphi}}(x,z,t)=0 \quad\mbox{and} \quad \lim_{|x|\rightarrow \infty}{ {\varphi}_1}(x,z,t)=0.
\end{equation}
The Euler equation  \eqref{E} can be expressed component-wise as 
\begin{equation*}
\begin{cases}
&\partial_t u + u u_{ x } + v u_{ z }+ 2 \omega v  = - \frac { P_{ x } } { \rho },   \\
&\partial_t v + u v_{ x } + v v_{ z } - 2 \omega u = -\frac { P_{ z } } { \rho } -g,
\end{cases}
\end{equation*}
in  the lower fluid layer $\Omega $, while in the upper fluid domain  $ \Omega_1$ the equations are 
\begin{equation*}
\begin{cases}
&\partial_t u_1 + u_1 u_{1,x} + v_1 u_{1,z}+ 2 \omega v_1  = - \frac { P_{1,x} } { \rho_1 }, \\
&\partial_t v_1 + u_1 v_{1,x} + v_1 v_{1,z} - 2 \omega u_1 = -\frac{ P_{1,z} }{ \rho_1 } -g.
\end{cases}
\end{equation*}
Using the expressions \eqref{def__stream_phi}, the Euler equations for the two layers can be written as
\begin{align*}
&\nabla \left( \varphi_t + \frac 1 2 |\nabla \psi|^2 - (\gamma+2\omega)\psi + \frac P {\rho} + gz  \right)=0\quad\text{in}\quad \Omega,\\
&\nabla \left( \varphi_{1,t} + \frac 1 2 |\nabla \psi_1|^2 - (\gamma_1+2\omega)\psi_1 + \frac {P_1} {\rho_1} + gz  \right)=0\quad\text{in}\quad \Omega_1,
\end{align*} where $\nabla=(\partial_x, \partial_z)^T$. 
The above system of Euler's equations written on the interface $z=\eta(x,t) $ where $P=P_1$ leads to a Bernoulli-type equation
\begin{align}
\rho\Big(({{\varphi}_{t}})_c&+\frac{1}{2}|\nabla \psi|_c^2-(\gamma+2\omega)\chi +g\eta\Big) \nonumber \\
&=\rho_1\Big(( {{\varphi}_{1,t}})_c+\frac{1}{2}|\nabla \psi_1|_c^2-(\gamma_1+2\omega)\chi +g\eta\Big)+ f(t) \label{Ber0}
\end{align} where $f(t)$ can be an arbitrary function and
\[
\chi(x,t)=- \int_{-\infty}^x\eta_t (x',t)dx'=-\partial_x^{-1}\eta_t
\]
is the stream function, evaluated at $z=\eta(x,t).$ However, enforcing decay conditions \eqref{sch},  also using \eqref{def__stream_phi}, we determine that $f$ is constant,
$f=\frac{1}{2}(\rho-\rho_1) \kappa^2,$ and the Bernoulli equation \eqref{Ber0} becomes
\begin{align}
    \rho (\varphi_t)_c&- \rho_1 (\varphi_{1,t})_c+\frac{\rho}{2}(\varphi_x^2+\varphi_z^2)_c-\frac{\rho_1}{2}(\varphi_{1,x}^2+\varphi_{1,z}^2)_c+[(\rho-\rho_1)g+(\rho \gamma -\rho_1 \gamma_1 )\kappa]\eta \nonumber \\
    &+\rho(\gamma \eta + \kappa) (\varphi_x)_c -\rho_1(\gamma_1 \eta + \kappa) (\varphi_{1,x})_c+\frac{\rho\gamma^2-\rho_1 \gamma_1^2}{2}\eta^2 - \Gamma \chi =0,  \label{Ber1}
\end{align}
where 
\begin{alignat}{2}
\Gamma:=\rho\gamma-\rho_1\gamma_1+2\omega\big(\rho-\rho_1\big)
\end{alignat}
is constant. 
As it stands, this expression  involves the two unknowns, $(\varphi)_c$ and $(\varphi_1)_c,$ however it will be shown that they can both be represented in terms of a single variable (together with $\eta$).

\end{section}

\section{Dirichlet-Neumann operators and evolution equations} \label{sec:DNO}
We introduce the following notations for the trace of the velocity potentials on $z=\eta(x,t)$:
\begin{equation}
        \left\lbrace
        \begin{array}{lcl}
        \phi:={\varphi}(x,\eta(x,t),t)\equiv (\varphi)_c,
      \qquad
        \phi_1:={\varphi}_1(x,\eta(x,t),t)\equiv (\varphi_1)_c.
        \end{array}
        \right.
\end{equation}
Given the existing assumptions on $\varphi,\varphi_1$ and $\eta$, it follows that the potentials $\phi,\phi_1 \in \mathcal{S}(\mathbb{R})$, and also the generalised momentum variable $\xi \in \mathcal{S}(\mathbb{R})$, for all $t$, where $\xi$ is defined \cite{BB} as
\begin{equation} \label{xi}
\xi:=\rho\phi-\rho_1\phi_1.
\end{equation}
 In order to  obtain evolution equations in terms of variables defined at the interface, we introduce the Dirichlet-Neumann (DN) operators \cite{Lan,CGK,CraigSulem,CGS}
\begin{equation}\label{DNOdef}
        \left\lbrace
        \begin{array}{lcl}
        G(\eta)\phi =({\varphi}_{{\bf{n}}})_c\sqrt{1+(\eta_x)^2} = ({\bf{n}}\cdot \nabla \varphi)_c\sqrt{1+(\eta_x)^2}, \quad \quad  \mbox{ for $\Omega$},
        \\
        G_1(\eta)\phi_1 =({\varphi}_{1,{\bf{n}}_1})_c\sqrt{1+(\eta_x)^2} = -({\bf{n}}\cdot \nabla \varphi_1)_c\sqrt{1+(\eta_x)^2}, \quad \quad \mbox{ for $\Omega_1$},
        \end{array}
        \right.
\end{equation}
where $\varphi_{{\bf{n}}}$ and $\varphi_{1,{\bf{n}}_1}$ are normal derivatives of the velocity potentials at the interface, and ${{\bf{n}}}=\frac{(-\eta_x, 1)}{\sqrt{1+(\eta_x)^2} }=-{{\bf{n}}_1}$ are outward unit normals  at the interface to the corresponding fluid domains.
We introduce also the operator
$B(\eta):=\rho G_1(\eta)+\rho_1 G(\eta)$.
Using the kinematic boundary conditions \eqref{KBC} and \eqref{DNOdef} we have
\begin{alignat}{2}\label{DNOetat}
        \left\lbrace
        \begin{array}{lcl}
        G(\eta)\phi=-\eta_x({\varphi}_x)_c+ ({\varphi}_z)_c = \eta_t+(\gamma\eta+\kappa)\eta_x,
        \\
        G_1(\eta)\phi_1=\eta_x({\varphi}_{1,x})_c-({\varphi}_{1,z})_c=-\eta_t-(\gamma_1\eta+\kappa_1)\eta_x
        \end{array}
        \right.
\end{alignat}
and adding these equations gives \begin{equation}\label{G1a}
    G(\eta)\phi+G_1(\eta)\phi_1=(\gamma-\gamma_1)\eta \eta_x.\end{equation} 
With \eqref{xi} and \eqref{G1a} we obtain
\begin{alignat}{2}\label{G2b}
B(\eta)\phi\equiv \rho_1G(\eta)\phi+\rho G_1(\eta)\phi=\rho_1 (\gamma-\gamma_1)\eta \eta_x + G_1(\eta)\xi.
\end{alignat}
We may now solve \eqref{G2b} for $\phi(x,t)$, and then use  \eqref{G1a} to find $\phi_1(x,t)$ in terms of $\xi$ and $\eta$: 
\begin{alignat}{2}\label{phys}
        \left\lbrace
        \begin{array}{lcl}
        \phi(x,t) =B^{-1}\big(\rho_1 (\gamma-\gamma_1)\eta \eta_x +G_1(\eta)\xi \big)= : \Phi[\xi(x,t), \eta(x,t)],
        \\
        \phi_1(x,t)  =B^{-1}\big(\rho (\gamma-\gamma_1)\eta \eta_x -G(\eta)\xi\big )=: \Phi_1[\xi(x,t), \eta(x,t)],
        \end{array}
        \right.
\end{alignat} where $\Phi, \Phi_1$ are functionals of $\xi$ and $\eta.$ 
From  \eqref{DNOetat} and  \eqref{phys} the time-evolution for $\eta$ can be finally expressed in terms of the DNO operators as
\begin{equation}
    \eta_t + \kappa \eta_x +[\gamma-\rho_1(\gamma-\gamma_1) G B^{-1}](\eta \eta_x) - GB^{-1}G_1 \xi =0,
\end{equation}
and also in an equivalent form as 
\begin{equation}
    \eta_t + \kappa \eta_x +[\gamma_1-\rho(\gamma_1-\gamma) G_1 B^{-1}](\eta \eta_x) - G_1B^{-1}G  \xi =0.
\end{equation}
It can be shown (see for example the expansions in \cite{CuIv}) that $G, G_1 $ and $B$, as well as the operator $GB^{-1}G_1$, are self-adjoint operators, thus 
$GB^{-1}G_1=G_1B^{-1}G$.
We have also
\begin{equation}\label{KBCDNO}
    \begin{split}
             &\eta_t + \kappa \eta_x +\gamma\eta \eta_x =\rho_1(\gamma-\gamma_1) G B^{-1}(\eta \eta_x) + GB^{-1}G_1 \xi =:F[\xi, \eta], \\
     &\eta_t + \kappa \eta_x +\gamma_1\eta \eta_x=\rho(\gamma_1-\gamma) G_1 B^{-1}(\eta \eta_x) +G_1B^{-1}G \xi =: F_1[\xi, \eta],
\end{split}
\end{equation}
which defines $F$ and $F_1$ as functionals of $\xi$ and $\eta.$
From \eqref{KBC} we have 
\begin{align}
     &\eta_t + \kappa \eta_x +\gamma\eta \eta_x =-\eta_x({\varphi}_x)_c+ ({\varphi}_z)_c  =F[\xi, \eta], \\
     &\eta_t + \kappa \eta_x +\gamma_1\eta \eta_x= -\eta_x({\varphi}_{1,x})_c+({\varphi}_{1,z})_c = F_1[\xi, \eta].
\end{align}
Noting that, from \eqref{phys}, 
\[ 
\phi_x = ({\varphi}_x)_c+ ({\varphi}_z)_c \eta_x = \Phi_x, 
\quad 
     \phi_{1,x} = ({\varphi}_{1,x})_c+ ({\varphi}_{1,z})_c \eta_x = \Phi_{1,x},
\]
we thus have a system for $({\varphi}_x)_c$  and $({\varphi}_z)_c$, given by
\[
    -\eta_x({\varphi}_x)_c+ ({\varphi}_z)_c  =F[\xi, \eta], \quad 
     ({\varphi}_x)_c+ ({\varphi}_z)_c \eta_x = \Phi_x[\xi, \eta],
\]
and for $(\varphi_{1,x})_c$  and $(\varphi_{1,z})_c$, given by
\[
    -\eta_x(\varphi_{1,x})_c+ (\varphi_{1,z})_c  =F_1[\xi, \eta], \quad
     ({\varphi}_{1,x})_c+ ({\varphi}_{1,z})_c \eta_x = \Phi_{1,x}[\xi, \eta],
\]
from which  these quantities can be expressed as functionals of $\xi$ and $\eta:$
\begin{align}\label{allphi}
    ({\varphi}_x)_c&=\frac{\Phi_x - \eta_x F }{1+\eta_x^2}, \quad ({\varphi}_z)_c=\frac{F+ \eta_x\Phi_x }{1+\eta_x^2}, \\
     ({\varphi}_{1,x})_c&=\frac{\Phi_{1,x} - \eta_x F_1 }{1+\eta_x^2}, \quad ({\varphi}_{1,z})_c=\frac{F_1+ \eta_x\Phi_{1,x} }{1+\eta_x^2}. 
\end{align}
Moreover, an expression for the $t$ derivatives comes from the chain-rule identity
$\phi_t=(\varphi_t)_c+(\varphi_z)_c \eta_t$, combined with the analogous identity for $\phi_1$, to give
\begin{align}\label{tder}
    \rho (\varphi_t)_c -  \rho_1 (\varphi_{1,t})_c &= ( \rho \phi-  \rho_1 \phi_1)_t -\eta_t [ \rho (\varphi_z)_c -  \rho_1 (\varphi_{1,z})_c  ]   \nonumber  \\
    &=\xi_t -\eta_t \frac{\rho F - \rho_1 F_1 + \eta_x \xi_x}{1+\eta_x^2}.
\end{align}
Using \eqref{allphi}--\eqref{tder} in  the Bernoulli type equation \eqref{Ber1} allows it to be recast  as a functional of $\xi$ and $\eta:$  
\begin{align}\label{Ber}
   & \xi_t -\eta_t \frac{\rho F - \rho_1 F_1 + \eta_x \xi_x}{1+\eta_x^2}+ \frac{\rho \Phi_x^2-\rho_1 \Phi_{1,x}^2+\rho F^2 - \rho_1 F_1^2}{2(1+\eta_x^2)}+ [(\rho-\rho_1)g+(\rho \gamma - \rho_1 \gamma_1) \kappa]\eta \nonumber \\
    &+\frac{\rho(\gamma \eta+\kappa)(\Phi_x-\eta_x F)}{1+ \eta_x^2}-\frac{\rho_1(\gamma_1 \eta+\kappa)(\Phi_{1,x}-\eta_x F_1)}{1+ \eta_x^2}+\frac{\rho\gamma^2-\rho_1 \gamma_1^2}{2}\eta^2 + \Gamma \partial_x ^{-1}\eta_t =0. \quad
\end{align}
Relations \eqref{KBCDNO} and \eqref{Ber} represent a closed system of evolution equations for $\eta$ and $\xi$ which involve only the DN operators $G(\eta)$ and $G_1(\eta).$ This system can be used to derive various approximate models  by expanding the DN operators according to the small parameters of the chosen propagation regime. However, it turns out to be preferable to work with the equivalent Hamiltonian formulation of the problem, as is outlined in the next section.

\section{Hamiltonian formulation} \label{sec:Ham}
The development of Hamiltonian formulations for various water wave problems was initiated by  the seminal work of Zakharov \cite{Zakharov}, which provided a framework for describing the propagation of a  water wave  on the surface of an infinitely deep fluid domain. For internal waves, a Hamiltonian approach utilising Dirichlet-Neumann operators was developed by Craig and collaborators in \cite{CGK,CraigSulem,CGS}. The creation of a Hamiltonian formulation which provided a model for internal waves propagating in the presence of  shear currents was more recently developed in  \cite{IC,CIT,CI,CIM,CuIv}. For such internal wave-current interactions, the Hamiltonian framework may be constructed as follows.

The total energy ($E$) of the system is a sum of the kinetic ($E_k$) and the potential ($E_p$) energies of the fluid flow, 
\begin{align}
E_k&=\frac{1}{2}\rho \int\limits_{\mathbb{R}} \int\limits_{-h}^{\eta(x,t)} (u^2+v^2)dz dx+\frac{1}{2}\rho_1 \int\limits_{\mathbb{R}} \int\limits_{\eta(x,t)}^{h_1} (u_1^2+v_1^2)dz dx,     \\
 E_p&=\frac{1}{2}\rho g \int\limits_{\mathbb{R}} \int\limits_{-h}^{\eta(x,t)} z\, dz dx +\frac{1}{2}\rho_1 g\int\limits_{\mathbb{R}} \int\limits_{\eta(x,t)}^{h_1} z\, dz dx.
\end{align}
The energy $E=E_k+E_p$ can be expressed as a functional of the system variables $\xi, \eta \in \mathcal{S}(\mathbb{R})$, with the resulting Hamiltonian then represented as a well-defined renormalised functional
\begin{equation}
    H(\xi, \eta)=E(\xi, \eta)-E(0,0),
\end{equation}
whose explicit form is given (see \cite{IC}) by
 \begin{align}
H  & =    \frac  1 2      \int_{ \R } \xi G( \eta ) B^{-1} G_1 ( \eta ) \xi \, dx +\frac 1 2 \left[ (\rho - \rho_1 ) g + ( \rho \gamma - \rho_1 \gamma_1 )\kappa \right]    \int_{ \R } \eta^2 dx   + \kappa \int_{ \R } \xi_x    \eta  dx  \nonumber \\
& + \rho_1 ( \gamma -\gamma_1 )  \int_{ \R } \eta  \eta_x  B^{-1} G ( \eta ) \xi dx - \frac 1 2 \rho \rho_1 ( \gamma - \gamma_1 )^2 \int_{ \R } \eta \eta_x B^{-1} \eta \eta_x dx   \nonumber \\
&  +  \frac 1 2   \gamma \int_{ \R } \xi_x  \eta^2  dx + \frac 1 6 \left( \rho \gamma^2 - \rho_1 \gamma_1^2 \right) \int_{\R} \eta^3 dx.  \label{HamDNO}
\end{align}
The equations of motion \eqref{KBCDNO} and \eqref{Ber} can be re-expressed (see details in  \cite{Compelli1Wavemotion,IC}) in terms of the (non-canonical) Hamiltonian form
\begin{equation}
\label{Hsys}
        \left\lbrace
        \begin{array}{lcl}
        \eta_t=\delta_{\xi} H,
        \\
        \xi_t=-\delta_{\eta} H+\Gamma  \chi,
        \end{array}
        \right.
\end{equation}
which can (formally) be re-expressed in the canonical Hamiltonian form
\begin{equation}
        \left\lbrace
        \begin{array}{lcl}
       \eta_t=\delta_{\zeta} H,
        \\
        \zeta_t=-\delta_{\eta} H
        \end{array}
        \right.
\end{equation}
under the change of  variables
$\xi\mapsto \zeta=\xi-\frac{\Gamma}{2} \int_{-\infty}^{x} \eta(x',t)\,dx'$.
\begin{remark}
In the formal limit $ \gamma,\gamma_1, \omega, \rho_1\to 0,$ the operator $ B^{-1} G_1$ corresponds to multiplication with $ \rho^{ -1 }$ and the Hamiltonian \eqref{HamDNO} reduces to
\[
 \frac {1} { 2 \rho } \int_{ \R } \xi G( \eta ) \xi dx + \frac 1 2 \rho g \int_{ \R }  \eta^2  dx,
\]
which matches the Hamiltonian for the one fluid layer irrotational  problem, see \cite{CraigSulem}.
\end{remark}

\section{Asymptotic expansion for the Hamiltonian}\label{sec:asympt}
In order to perform an asymptotic analysis we introduce the necessary nondimensionalised parameters. Let $ \lambda $ denote a characteristic wavelength for the internal waves that we consider, while $\eta_{\text{max}}$ denotes a maximal permitted wave amplitude. 
Define the steepness, and shallowness, parameters by
\[
\eps = \frac{|\eta_{\text{max}}|}{h},\quad \delta = \frac{h}{\lambda},
\]  respectively  \cite{CJ08}. We assume that $h$ and $h_1$ are of the same order, $h=\mathcal O(h_1)$ in terms of Landau notation.
Initially, we will focus on the small amplitude regime. The DNO for the lower fluid domain $\Omega$ allows an asymptotic expansion of the form \cite{CGK,CraigSulem,CGS}
$$ G( \eta )=  G^{ (0 ) } +G^{ (1 ) } ( \eta )+G^{ (2 ) } ( \eta )\ldots ,$$ 
where $G^{ (n ) } ( \eta )$ is a homogeneous expression of $\eta$ of order $n,$ that is, for any constant $\nu,$
$G^{ (n ) } ( \nu \eta )=\nu^n G^{ (n ) } ( \eta ).$ Thus, $G^{ (n ) } ( \eta )$ is of order $\varepsilon^n.$
The zeroth-order approximation of $ G( \eta ) $ is, for example,  $$ G^{ (0 ) } := D \tanh ( h D)$$ 
where $D=-i\partial_x$ and its action on on a single mode of the form $ e^{ i k x }$ is given by its Fourier multiplier $ k \tanh ( h k).$  
The next order approximation of $ G $ takes into account the contribution from the order $\eps$ term,
\begin{align*}
    G^{ (1 ) } ( \eta ) = D \eta D - G^{ (0 ) } \eta G^{ (0 ) },
\end{align*}
etc. Note  that the operators appearing in the above definitions are non-commutative.  
For the DNO of the upper layer, $ G_1(\eta),$ we  similarly have
\begin{align*}
G^{ (0 ) }_1 & = D \tanh ( h_1 D), \\
    G^{ (1 ) }_1 ( \eta )& =- D \eta D + G^{ (0 ) }_1 \eta G^{ (0 ) }_1,
\end{align*}
and so on, see \cite{CGK}.
In order to keep track of the order of the terms we replace $\eta \mapsto \eps \eta$ (following which, $\eta$ will be treated as an order one quantity). Moreover, we need to expand over the long-wave (shallowness) parameter $\delta$ as well. In view of the definitions of the shallowness parameter and the wave-number, we have $k h_1 =\mathcal{O} (\delta).$ Since $h,h_1$ are of the same order, we also have  $k h_1 =\mathcal{O} (\delta).$
Again, to keep track of the orders, we substitute $ h D \mapsto \delta \, h D$,  $ h_1 D \mapsto \delta \, h_1 D$ (where all quantities have order one after scaling). A Taylor expansion, up to an appropriate order, yields   
$$ \tanh D h =  \delta D h -   \delta^3 \frac{ ( D h )^3 } { 3 } +  \delta^5 \frac { 2 ( D h )^5 } { 15 } + \mathcal{ O } ( \delta^7  ).$$
Thus, we obtain
\begin{equation}\label{G0}
G( \eta ) =   \delta^2 \left( D^2 h  + \eps  D \eta D  \right) -   \delta^4 \left( \frac{  D^4 h^3 } { 3 }  + \eps   D^2 \eta D^2 h^2 \right)+ \delta^6  \frac { 2  D^6 h^5 } { 15 }  + \mathcal{ O } ( \delta^8),
\end{equation}
and
\begin{equation}\label{G1}
G_1( \eta ) =   \delta^2 \left( D^2 h_1  - \eps  D \eta D  \right) +   \delta^4 \left( -\frac{  D^4 h_1^3 } { 3 }  + \eps   D^2 \eta D^2 h_1^2 \right)+ \delta^6  \frac { 2  D^6 h_1^5 } { 15 }  + \mathcal{ O } ( \delta^8).
\end{equation}
Now scaling the $\xi-$variable by $\xi \mapsto \delta \, \xi$  ensures that a balance occurs between the small amplitude, $\eps$, and shallowness, $\delta$, parameters of the form $\eps\sim \delta^2$: this balance between parameters is known as the KdV propagation regime. The rationale behind this choice of scaling is  that the kinetic energy density has the same order as $\xi_x^2/2$ (since $\xi_x$ is a horizontal velocity), and the order of the kinetic energy must be matched by that of the potential energy density, given by $\eta^2/2$, which implies that $\xi_x \sim \eta$. In terms of scaled variables, this relation becomes $\delta^2 \xi_x\sim \eps \eta$ (since the previous scaling of $\xi$ leads to  $\xi_x \mapsto \delta^2 \xi_x$, bearing in mind that the $x-$derivative has the same order as $D$, which is scaled by $\delta$) which further  implies that  $\eps\sim \delta^2$.

Using expansions \eqref{G0} and \eqref{G1}  in the Hamiltonian expression \eqref{HamDNO}, applying the above scaling for $\xi$, and retaining terms up to order $\delta^8$, we have 
\begin{multline}\label{Ham8}
H(\eta,\xi)=\frac{\delta^4}{2}\ii \xi
 \Bigg(
\frac{h h_1}{\rho_1 h+\rho h_1}{D}^2
+\varepsilon\frac{\rho h_1^2-\rho_1 h^2}{(\rho_1 h+\rho h_1)^2}
D \eta  D -\varepsilon^2  \frac{\rho \rho_1(h+h_1)^2}{(\rho_1 h+\rho h_1)^3}D \eta^2  D\Bigg)\xi  dx\\
+\frac{\delta^6}{2}\ii  \xi
 \Bigg(-\frac{h^2 h_1^2(\rho_1   h_1+\rho h)}{3(\rho_1 h+\rho h_1)^2}{D}^4
-\varepsilon \frac{h^2 h_1^2(\rho-\rho_1)}{(\rho_1 h+\rho h_1)^2} {D}^2\eta  {D}^2\\
+ \varepsilon \frac{\rho \rho_1 h h_1(h- h_1)(h+ h_1)^2}{3(\rho_1 h+\rho h_1)^3}(D \eta D^3+D^3 \eta D)
\Bigg)\xi  dx\\
+\frac{\delta^8}{2}\ii \xi
 \frac{ h^2 h_1^2\left( \rho \rho_1 (h^2-h_1^2)^2+6 h h_1  (\rho h+\rho_1 h_1)^2\right)}{45(\rho_1 h+\rho h_1)^3} D^6
\xi  dx\\
-\varepsilon\delta^2 \kappa\ii \xi\eta_x \,dx+\varepsilon^2\delta^2\frac{\rho_1\gamma_1 h + \rho \gamma h_1}{2(\rho_1 h+\rho h_1)}\ii\eta^2 \xi_x\,dx
\\
-\varepsilon^3\delta^2\frac{\rho \rho_1 (\gamma-\gamma_1)(h+h_1)}{2(\rho_1 h+\rho h_1)^2}\ii\eta^{3}\xi_x \,dx
-\varepsilon^2 \delta^4\frac{\rho \rho_1h h_1(\gamma-\gamma_1)(h^2-h_1^2)}{6(\rho_1 h+\rho h_1)^2}\ii\eta^{2}\xi_{xxx}\,dx\\
+\frac{\varepsilon^2}{2}[g(\rho-\rho_1 )+(\rho\gamma-\rho_1\gamma_1)\kappa]\ii  \eta^2 dx +\frac{\varepsilon^3}{6}(\rho\gamma^2-\rho_1\gamma_1^2)\ii  \eta^3dx \\
- \frac{\varepsilon^4}{2}\frac{\rho\rho_1(\gamma-\gamma_1)^2}{\rho_1 h+\rho h_1}\ii   \frac{\eta^4}{4} \,dx.
\end{multline}
Further details concerning the derivation of \eqref{Ham8} can be found  in \cite{IC}.

\section{Higher order KdV type equation} \label{sec:HKdV}
\subsection{Hamiltonian and system of higher-order coupled equations}
As it is more convenient to work with the variable  $ \mathfrak{u}=\xi_x  $ rather than  $\xi$,  equations \eqref{Hsys} can be rewritten in terms of $\mathfrak{u} $ and $\eta$ as
 \begin{equation} \label{H_u_eta}
\eta_t =-\Big(\frac{\delta H}{\delta \mathfrak{u}}\Big)_x \quad \text{and} \quad \mathfrak{u}_t=-\Big(\frac{\delta H}{\delta \eta}\Big)_x-\Gamma\eta_t,
\end{equation}
for which the  Hamiltonian  \eqref{Ham8} is recast as
\begin{multline} \label{H}
H(\eta,\mathfrak{u})= \varepsilon^2 \frac{1}{2}\alpha_1 \int_{\mathbb{R}}  \mathfrak{u}^2 \, dx 
+\varepsilon^2 \frac{1}{2}\alpha_5 \int_{\mathbb{R}}  \eta^2 \, dx   
+\varepsilon^2 \kappa \int_{\mathbb{R}}\eta  \mathfrak{u} \, dx  \\
-\varepsilon^3 \frac{1}{2}\alpha_2 \int_{\mathbb{R}}  \mathfrak{u}_x^2 \, dx
+ \varepsilon^3 \frac{1}{2}\alpha_3 \int_{\mathbb{R}}\eta  \mathfrak{u}^2 \, dx 
+\varepsilon^3 \frac{1}{2}\alpha_4 \int_{\mathbb{R}}\eta^2  \mathfrak{u} \, dx 
+ \varepsilon^3 \alpha_6 \int_{\mathbb{R}}\frac{\eta^3}{6}  \, dx\\
-\varepsilon ^4 \frac{1}{2} \beta_1 \int_{\mathbb{R}}  \eta^2 \mathfrak{u} ^2 \,dx
-\varepsilon ^4 \frac{1}{2} \beta_2 \int_{\mathbb{R}}  \eta \mathfrak{u}_x ^2 \,dx
-\varepsilon ^4  \beta_3 \int_{\mathbb{R}}  \eta \mathfrak{u} \mathfrak{u}_{xx}\, dx
+\varepsilon ^4 \frac{1}{2} \beta_4 \int_{\mathbb{R}}   \mathfrak{u}_{xx} ^2 \,dx
\\
-\varepsilon ^4  \beta_5 \int_{\mathbb{R}}  \eta^3 \mathfrak{u} \, dx
-\varepsilon ^4  \beta_6 \int_{\mathbb{R}}  \eta^2 \mathfrak{u}_{xx} \, dx
-\varepsilon ^4 \frac{1}{2} \beta_7 \int_{\mathbb{R}}  \frac{\eta^4}{4} \, dx.
\end{multline}
Here the parameters are defined as 
\begin{equation}\label{HamC} \begin{split} \alpha_1&=\frac{h h_1}{\rho_1 h+\rho h_1}, \qquad \alpha_2= \frac{h^2h_1^2(\rho h+\rho_1 h_1)}{3(\rho_1 h+\rho h_1)^2}, \qquad \alpha_3= \frac{\rho h_1^2-\rho_1 h^2}{(\rho_1 h+\rho h_1)^2}, \\
\alpha_4&=\frac{\gamma_1\rho_1 h+ \gamma \rho h_1}{\rho_1 h+\rho h_1}, \quad \alpha_5= g(\rho-\rho_1)+(\rho \gamma -\rho_1 \gamma_1)\kappa, \quad \alpha_6=\rho\gamma^2-\rho_1 \gamma_1^2,
\end{split}\end{equation}
 and \begin{align}\beta_1&=\frac{\rho \rho_1(h+h_1)^2}{(\rho_1 h + \rho h_1)^3}, \quad \beta_2=\frac{h^2 h_1^2(\rho-\rho_1 )}{(\rho_1 h + \rho h_1)^2} , \quad \beta_3= \frac{h h_1 \rho \rho_1(h-h_1 )(h+h_1)^2}{3(\rho_1 h + \rho h_1)^3}, \nonumber\\
 \beta_4&= \frac{h^2 h_1^2[ \rho \rho_1(h^2-h_1^2 )^2+6hh_1(\rho h+\rho_1h_1)^2]}{45(\rho_1 h + \rho h_1)^3}, \nonumber \\
 \beta_5 &=\frac{\rho \rho_1 (\gamma-\gamma_1)(h+h_1)}{2(\rho_1 h+\rho h_1)^2}, \quad \beta_6=\frac{\rho \rho_1h h_1(\gamma-\gamma_1)(h^2-h_1^2)}{6(\rho_1 h+\rho h_1)^2}, \nonumber \\
 \beta_7 &=\frac{\rho\rho_1(\gamma-\gamma_1)^2}{\rho_1 h+\rho h_1}.
  \end{align}
The Hamiltonian equations \eqref{H} can be expressed in terms of $\eta$ and $\mathfrak{u}$ as
\begin{align}\label{BA0}
\eta_t  =&-\left( \frac{\delta H}{\delta \mathfrak{u}} \right)_x \nonumber \\
 = &-[ \alpha_1 \mathfrak{u} +\kappa \eta + \varepsilon \alpha_2 \mathfrak{u}_{xx} + \varepsilon \alpha_3 \eta \mathfrak{u}+\varepsilon \alpha_4 (\eta^2)/2 - \varepsilon ^2 \beta_1 \eta^2 \mathfrak{u}+ \varepsilon^2 \beta_2 (\eta \mathfrak{u}_x)_x \nonumber \\
 &  - \varepsilon ^2 \beta_3 \eta \mathfrak{u}_{xx} - \varepsilon^2 \beta_3 (\eta \mathfrak{u})_{xx}+  \varepsilon^2 \beta_4 \mathfrak{u}_{xxxx}- \varepsilon^2 \beta_5 \eta^3 - \varepsilon^2 \beta_6 (\eta^2)_{xx}]_x, \nonumber \\
 \mathfrak{u}_t+\Gamma \eta_t &= -\left( \frac{\delta H}{\delta \eta} \right)_x  \nonumber \\
=& - [\alpha_5 \eta + \kappa \mathfrak{u}  + \varepsilon \alpha_3 (\mathfrak{u}^2) / 2 + \varepsilon \alpha_4 \eta \mathfrak{u} + \varepsilon \alpha_6 (\eta ^2)/2  -\varepsilon^2 \beta_1 \eta \mathfrak{u}^2 -\varepsilon^2 \beta_2 (\mathfrak{u}_x)^2/2         \nonumber \\
&-\varepsilon^2 \beta_3 \mathfrak{u}\mathfrak{u}_{xx} -\varepsilon^2 \beta_5 3 \eta^2 \mathfrak{u}-\varepsilon^2 \beta_6 2 \eta \mathfrak{u}_{xx}
-\varepsilon^2 \beta_7 (\eta^3)/2]_x,
\end{align}
where $ \Gamma:= \rho \gamma - \rho_1 \gamma_1 + 2 \omega (\rho - \rho_1)$.
Note that, when $ \Gamma \neq 0 $, \eqref{BA0} is  a nearly-Hamiltonian system (similar to \cite{CIP}).
The Galilean change of variables $x \mapsto x-\kappa t$ (corresponding to a move to the reference frame travelling with speed $\kappa$) leads to the disappearance of all terms involving $\kappa$, since $\partial_t+\kappa \partial_x  \mapsto  \partial_t$.  
In what follows, we assume that $ g \gg \omega \kappa$, which is reasonable since  $\omega= 7.3\times 10^{-5}$ rad/s and $ \kappa$ (the mean underlying current speed) is usually not excessively large (typical maximal speeds are of the order of several metres per second) . Thus we may neglect the term $\omega \kappa.$ Replacing  $\alpha_5$, via \eqref{HamC}, we rewrite \eqref{BA0} as 
\begin{align}
\eta_t  = &-[ \alpha_1 \mathfrak{u}_x + \varepsilon \alpha_2 \mathfrak{u}_{xxx} + \varepsilon \alpha_3 (\eta \mathfrak{u})_x+\varepsilon \alpha_4 \eta \eta_x - \varepsilon ^2 \beta_1 (\eta^2 \mathfrak{u})_x+ \varepsilon^2 \beta_2 (\eta \mathfrak{u}_x)_{xx}  - \varepsilon ^2 \beta_3 (\eta \mathfrak{u}_{xx})_x 
\nonumber \\
 & - \varepsilon^2 \beta_3 (\eta \mathfrak{u})_{xxx}+  \varepsilon^2 \beta_4 \mathfrak{u}_{xxxxx}- \varepsilon^2 \beta_5 (\eta^3)_x - \varepsilon^2 \beta_6 (\eta^2)_{xxx}] := f_1, \label{BA2Eta} \\
 \mathfrak{u}_t+ &\Gamma \eta_t =  \mathfrak{u}_t+\Gamma f_1 \nonumber \\
=& -[(\rho-\rho_1)g  \eta_x   + \varepsilon \alpha_3 \mathfrak{u} \mathfrak{u}_x + \varepsilon \alpha_4 (\eta \mathfrak{u})_x + \varepsilon \alpha_6 \eta \eta_x  -\varepsilon^2 \beta_1 (\eta \mathfrak{u}^2)_x -\varepsilon^2 \beta_2 \mathfrak{u}_x \mathfrak{u}_{xx}         \nonumber \\
&-\varepsilon^2 \beta_3 (\mathfrak{u}\mathfrak{u}_{xx})_x -\varepsilon^2 \beta_5 3 (\eta^2 \mathfrak{u})_x-\varepsilon^2 \beta_6 2( \eta \mathfrak{u}_{xx})_x -\varepsilon^2 \beta_7 (\eta^3)_x/2] :=f_2+\Gamma f_1.\label{BA2}
\end{align}
Substituting the expression for $\eta_t$ from the first equation into the second gives
\begin{align}\label{BA3}
 & \mathfrak{u}_t  =   
  \Gamma [ \alpha_1 \mathfrak{u}_x + \varepsilon \alpha_2 \mathfrak{u}_{xxx} + \varepsilon^2 \beta_2 (\eta \mathfrak{u}_x)_{xx}   - \varepsilon^2 \beta_3 (\eta \mathfrak{u})_{xxx}+  \varepsilon^2 \beta_4 \mathfrak{u}_{xxxxx}- \varepsilon^2 \beta_6 (\eta^2)_{xxx}] \nonumber \\
& -[(\rho-\rho_1)g  \eta_x   + \varepsilon \alpha_3 \mathfrak{u} \mathfrak{u}_x + \varepsilon ( \alpha_4   -\Gamma \alpha_3 ) (\eta \mathfrak{u})_x + \varepsilon ( \alpha_6 - \Gamma \alpha_4 ) \eta \eta_x  -\varepsilon^2 \beta_1 (\eta \mathfrak{u}^2)_x \nonumber \\
& -\varepsilon^2 \beta_2 \mathfrak{u}_x \mathfrak{u}_{xx}-\varepsilon^2 \beta_3 (\mathfrak{u}\mathfrak{u}_{xx})_x -\varepsilon^2 ( 3 \beta_5   - \Gamma \beta_1) (\eta^2 \mathfrak{u})_x-\varepsilon^2 ( 2 \beta_6    - \Gamma \beta_3 )(\eta \mathfrak{u}_{xx})_x \nonumber \\
& -\varepsilon^2 ( \beta_7/2 - \Gamma \beta_5 ) (\eta^3)_x ].
\end{align}

\subsection{Reduction to a single higher-order equation for $\eta$}
To gain insight into equations   \eqref{BA2Eta} and \eqref{BA3}, we examine their linearisation by way of invoking the plane-wave ansatz  $ \eta = \eta_0 e^{ i k(x-c t)}$ and $ \u = \u_0 e^{ i k(x-c t)}$, where $c$ represents the wave-speed in a frame moving with speed $\kappa.$  The zero order approximation ($\varepsilon=0$) is given by
\begin{equation} \label{linearLevel}    
 \mathfrak{u}_t  =   
  \Gamma  \alpha_1 \mathfrak{u}_x - (\rho-\rho_1)g  \eta_x \quad \eta_t  = - \alpha_1 \mathfrak{u}_x, 
\end{equation}
which results in the following dispersion relation for the wave speed:
\begin{equation}\label{wavespeed}
c= \frac{1}{2}\left( - \alpha_1 \Gamma \pm \sqrt{ \alpha_1 ^2 \Gamma^2 + 4 \alpha_1 ( \rho -\rho_1 ) g }\right ),    
\end{equation} where the plus sign represents right-running waves and the minus sign leads to left-running waves. 
The expression \eqref{wavespeed} also appears in \cite{IC,CIT}. Note that, at the linear level, from \eqref{linearLevel} we have $\eta = \frac{\alpha_1}{c} \u.$
In the next order of approximation, one can show that $\eta$ will satisfy the KdV equation \cite{KDV}. In order to proceed to a higher level of approximation, we make an Ansatz of the form 
\begin{equation}\label{ansatz}
\u=A_1 \eta+\eps \left( B_1 \eta^2 + B_2 \eta_{xx}\right) +\eps^2 \left( C_1 \eta^3 + C_2 \eta \eta_{xx}+ C_3 \eta^2_x + C_4  \eta_{xxxx} \right):= \L (\eta),
\end{equation}
where $A_1=\frac{c}{\alpha_1}$, with as yet unknown coefficients that will be determined below.
Our aim now is to use \eqref{ansatz} to find asymptotic expansions for $f_1$ and $f_2$ (as defined by \eqref{BA2Eta} and \eqref{BA2}) which can be expressed in terms of $\eta$ alone. Then the two equations $\eta_t=f_1[\eta]$ and $\u_t=f_2[\eta]$ are compatible if and only if 
\begin{equation}\label{compat}
     f_2[\eta] = \delta \L ( \eta ) f_1[\eta]
     \end{equation}
     where the operator $\delta \L ( \eta ) $ is given by the variational derivative of $\L$, that is, for a sufficiently smooth function $f$:
\begin{equation}\label{ansatz1}
\delta \L( \eta ) f = A_1 f +  \eps \left( 2 B_1 \eta f +  B_2   f_{xx} \right)  + \eps \left( 3 C_1 \eta^2 f + C_2 \eta_{xx}f  + C_2 \eta f_{xx}  + 2 C_3 \eta_{x}  f_x + C_4 f_{xxxx}  \right).
\end{equation}
Indeed, one can check that from \eqref{ansatz} it easily follows that $ \u_t = \delta \L ( \eta ) \eta_t$.
From \eqref{BA2} and \eqref{ansatz} we obtain 
\begin{align}\label{AUX1a}
 -f_1[\eta] & =  \alpha_1 A_1 \eta_x  + \eps  \left[  ( \alpha_1 B_2 + \alpha_2 A_1 ) \eta_{xxx} + (  2 \alpha_1 B_1 +  2 \alpha_3 A_1 + \alpha_4 ) \eta \eta_x  \right] \nonumber \\
& + \eps^2   \left[ 3 ( C_1 \alpha_1 +  \alpha_3 B_1 -  \beta_1  A_1 -  \beta_5 ) \eta^2 \eta_x \right. \nonumber \\
&\left. + ( \alpha_1 C_2 + 2 \alpha_1 C_3 + 6 \alpha_2  B_1  +  \alpha_3  B_2 + 3 \beta_2 A_1 - 7 \beta_3 A_1 - 6 \beta_6 )\eta_x \eta_{xx} \right. \nonumber  \\
&\left. + ( \alpha_1 C_2 + 2 \alpha_2 B_1 + \alpha_3 B_2 + \beta_2 A_1 - 3 \beta_3 A_1 - 2 \beta_6    )\eta  \eta_{xxx} \right. \nonumber \\
&\left. + ( \alpha_1 C_4  +  \alpha_2 B_2   + \beta_4 A_1 )  \eta_{xxxxx} \right],
\end{align}
\begin{align}\label{AUX2b}
&-\left(f_2[\eta]  + \Gamma  f_1[\eta] \right)  = (\rho - \rho_1 ) g \eta_x \nonumber +  \eps\left(  \alpha_3  A_1^2   + 2 \alpha_4 A_1 + \alpha_6  \right) \eta \eta_x \nonumber\\
& + \eps^2\left[ \alpha_3 A_1  \left(  3 B_1  \eta^2 \eta_x  + B_2 \eta \eta_{xxx }  + B_2 \eta_x  \eta_{xx}\right) +  \alpha_4    \left(  3 B_1 \eta^2 \eta_{x} +  B_2 \eta_{x} \eta_{xx}  + B_2 \eta  \eta_{xxx}\right)\right.\nonumber \\
&\left.
 - 3  \beta_1  A_1^2     \eta^2 \eta_{x}   
 -  \beta_2  A_1^2     \eta_{x} \eta_{xx}    -   \beta_3  A_1^2 \left(    \eta_{x} \eta_{xx} + \eta \eta_{xxx}   \right)
 - 9  \beta_5  A_1    \eta^2  \eta_{x}\right. \nonumber\\
&\left.     - 2  \beta_6  A_1 \left(    \eta_x \eta_{xx} + \eta \eta_{xxx} \right)     -  3   \beta_7   \eta^2  \eta_{x} /2 \right],
\end{align}
thereby expressing $f_1$ and $f_{2}$ as functions of $\eta$.   Hence we can infer that  $\eta$ satisfies an evolution equation of the form
\begin{equation}\label{etaf1}
    \eta_t=f_1[\eta],
    \end{equation} that is,
\begin{align}\label{KdV5}
 \eta_t + c \eta_x  + \varepsilon (  \tilde A   \eta_{xxx} +  A   \eta \eta_x)  + \varepsilon ^2 \!\!\left\{   M \eta^2 \eta_{ x } + N_1 \eta_{ x } \eta_{ x x } +  N_2 \eta \eta_{ xxx }  + Q \eta_{ xxxxx } \right\}  = \mathcal{O}(\varepsilon^3), \quad \ 
\end{align}
for some constant coefficients, which are yet to be determined. The method that we will follow, in order to do so, consists of an appropriate matching between the coefficients of the respective $\eta$-terms that occur in \eqref{compat}. 
We refer to the Appendix where the coefficients $A,$ $\tilde{A},$ $M,$ $N_1,$ $N_2$ and $Q,$ as well as  $A_1,$ $B_1,$ $B_2,$  $C_1,\ldots,C_4$ are derived in terms of the constant parameters of the system. As a result, the coefficients of equation \eqref{KdV5} are 
\begin{align}
A   &  = \frac { 3 \alpha_3  c^2 + 3 \alpha_1  \alpha_4 c + \alpha_6 \alpha_1^2 } { \alpha_1 ( 2 c + \Gamma \alpha_1 ) }   \nonumber \\
& = \frac {  \alpha_1   } { h^2 h_1^2  ( 2 c + \Gamma \alpha_1 )   }\left\{   \rho  h_1^2  \left[    \frac {  3 } { 4 }  c^2  + (  \frac {   3 } { 2 } c + \gamma  h )^2 \right]   - \rho_1 h^2 \left[  \frac {  3 } { 4 }   c^2 + (  \frac {   3 } { 2 }  c - \gamma_1 h_1 )^2 \right]     \right\},
\label{A}  \end{align}
\begin{equation}
    \tilde{A}  =  \frac {  \alpha_2 c^2 } { \alpha_1 ( 2 c + \Gamma \alpha_1 ) } 
=  \frac 1 3   \frac {  (  \rho h +  \rho_1 h_1 )  c^2 } {  ( 2 c + \Gamma \alpha_1 ) } \alpha_1, 
\label{Atilde}
\end{equation}
\begin{align}
 M = 3\left [    \alpha_1   B_1 \frac { 2 \alpha_3 A_1 + \alpha_4   - 4  \alpha_1 B_1  } { 3 ( 2 c + \Gamma \alpha_1 ) }  -  \frac { 2 \beta_1  A_1 c } { 2 c + \Gamma \alpha_1}  - \frac { 4 \beta_5 c } { 2 c + \Gamma \alpha_1}  - \frac {  \beta_7 \alpha_1  } { 2 ( 2 c + \Gamma \alpha_1 ) }  \right],
\label{M} \end{align}
\begin{align}
 N_1 = \frac {2   \beta_2   c^2   -   8  \beta_3  c^2   -   8 \alpha_1  \beta_6 c      +  8 \alpha_1   \alpha_2  B_1 c   - 2 \alpha_1^2 \left(    B_1 \tilde A  +  B_2 A   \right ) }   { \alpha_1 ( 2 c + \Gamma \alpha_1 ) }, 
\label{N1} \end{align}
\begin{align}
 N_2 &= -   \frac { 4 \alpha_1^2 B_1 B_2 } {  2 c + \Gamma \alpha_1  } +  \frac {  \beta_2  A_1 c } { 2 c + \Gamma \alpha_1}  -  \frac { 4 \beta_3  A_1 c } { 2 c + \Gamma \alpha_1}  - \frac { 4 \beta_6 c } { 2 c + \Gamma \alpha_1},
\label{N2} \end{align}
 \begin{align}
  Q:= \frac {  \beta_4  c^2  -  \alpha_1^3 B_2^2 } {  \alpha_1 ( 2 c + \Gamma \alpha_1 )  }, \label{Q} \end{align}
  where
 \begin{align*}
A_1 &=  \frac c { \alpha_1}, \nonumber \\
B_1 &=  -  \frac {  \alpha_3  c  }{   \alpha_1^2 } -   \frac {   \alpha_4    }{ 2  \alpha_1 } + \frac { 3 \alpha_3  c^2 + 3 \alpha_1  \alpha_4 c + \alpha_6 \alpha_1^2 } { 2 \alpha_1^2 ( 2 c + \Gamma \alpha_1 ) },
\nonumber \\
B_2 &=  \frac { -  \alpha_2  c   (  c + \Gamma \alpha_1  ) }{  \alpha_1^2 ( 2 c +  \Gamma  \alpha_1 )} = - \frac {   \alpha_2  c    }{  \alpha_1^2 } + \frac {   \alpha_2  c^2   }{  \alpha_1^2 ( 2 c +  \Gamma  \alpha_1 ) }.
\end{align*}
are the first three coefficients in \eqref{ansatz} (the others are given in the Appendix).
  
\begin{remark}
If we consider the right-running wave, for which $ 2 c + \Gamma \alpha_1>0$, then \eqref{Q} can be recast in the form
     \[ Q = \frac {  \rho \rho_1 ( h^4 + h_1^4 ) \alpha_1^2 c^2   } { 45 h h_1   ( 2 c + \Gamma \alpha_1 )  } + \frac {   ( \rho^2  h^2 + \rho_1^2  h_1^2 )  \alpha_1^2  c^2  } {  45   ( 2 c + \Gamma \alpha_1 )  }   + \frac {   ( \rho h + \rho_1 h_1 )^2  \alpha_1^2  c^2  } {  9  ( 2 c + \Gamma \alpha_1 )  } \left[    1  -   \left(    \frac {    c + \Gamma  \alpha_1 }{   2 c +  \Gamma  \alpha_1  } \right)^2    \right]. \]
In view of \eqref{wavespeed}, it follows that the coefficient of the higher-order dispersive term is positive.
\end{remark}
 \begin{remark}
     In the absence of currents and Coriolis force, we have the following expressions for the higher order coefficients in \eqref{KdV5}:
\begin{multline*}
M = - \frac {3  c } { \alpha_1 } \left(  \frac  {  \alpha_3^2 } { 8 \alpha_1 }  + \beta_1 \right),\quad    N_2 =  - \frac { c } { \alpha_1 }  \left(  \frac {  \alpha_2 \alpha_3 } { 4 \alpha_1 }- \beta_2 / 2 + 2 \beta_3 \right), \\
N_1 = \frac { c } { \alpha_1 } \left(    \beta_2 - 4 \beta_3    - \frac 1 8 \frac { \alpha_2 \alpha_3 } {  \alpha_1 }  \right), \quad    Q = \frac { \rho \rho_1 ( h^2 - h_1^2 )^2 } { 90 h h_1 } \alpha_1^2 c + \frac {19} {40}  \frac {  \alpha_2^2 c } { \alpha_1^2 }.
\end{multline*} 
 \end{remark}
 \begin{remark}
From \eqref{A} we notice that the situation with $A=0$ is possible for some values of the parameters. 
Let us consider again the right-running wave, when $ 2 c + \Gamma \alpha_1>0.$ If $  A  = 0  $, we want to further examine the sign of $ M $. In order to do so, we rewrite 
\begin{align*}
&4    ( h + h_1 )^2   c^2  + 4 h h_1 ( \gamma - \gamma_1 ) ( h + h_1 )  c  +   h^2 h_1^2 ( \gamma - \gamma_1 )^2\\
&  =   3   ( h + h_1 )^2 c^2 + 2 h h_1 ( \gamma - \gamma_1 ) ( h + h_1 ) +  \left(      ( h + h_1 ) c   +      h h_1 ( \gamma - \gamma_1 )  \right) ^2.
\end{align*}
Using the latter, we can easily deduce that $  M $ reduces to  
\begin{align*}
&-3 \alpha_1^2 \left [ \frac {    4  h^3 h_1^3   B_1^2  + \rho \rho_1\left[  3   ( h + h_1 )^2 c^2 + 2 h h_1 ( \gamma - \gamma_1 ) ( h + h_1 ) +  \left(      ( h + h_1 ) c   +      h h_1 ( \gamma - \gamma_1 )  \right) ^2 \right]    } { 2  h^3 h_1^3 ( 2 c + \Gamma \alpha_1 )}\right].
\end{align*}
 A direct implication of this expression is that $M<0$ when the coefficient of the term $\eta \eta_x$ vanishes and  $ \gamma \geq \gamma_1.$ In this case the  $M\eta^2\eta_x$ term balances the dispersive terms. The scaling then is slightly different, $\varepsilon \sim \delta,$  and the equation for $\eta$ acquires the form of the integrable Gardner equation $$\eta_t + c \eta_x + \varepsilon^2(\tilde{A} \eta_{xxx}+M\eta^2\eta_x )+\ldots=0.$$ Since $\tilde{A}>0,$ and $M<0$ we have a defocusing Gardner's equation or, in this case, a defocusing mKdV equation,
 $$\eta_t + c \eta_x + \tilde{A} \eta_{xxx}+M\eta^2\eta_x =0$$ (written without the scaling). This equation does not support soliton solutions, but instead
 there is a ``shock-wave'' solution of the form
\begin{equation}\label{sw}
    \eta(x,t)= \pm 2 K \sqrt{-\frac{3\tilde{A}}{M}}\tanh[\sqrt{2}K(x-x_0 - (c+4\tilde{A}K^2)t)],
\end{equation}  where $K$ and $x_0$ are constants (the soliton parameters).
\end{remark}

\section{ Asymptotic integrability of HKdV equation} \label{sec:Int}

In this section we will examine the so-called asymptotic integrability \cite{FL,KOD} of equation \eqref{KdV5}. An equation is called asymptotically integrable if it can be mapped, up to a certain degree of asymptotic accuracy, to an integrable one, see \cite{HK} and the references therein. In our case, we are interested in relating the (not necessarily integrable) equation \eqref{KdV5} to some integrable equations (cf. \cite{DGH1,DGH2,FL,KOD,KO,ZS} for examples in this direction for KdV-type equations related to water waves).
One can use the approach outlined in \cite{DGH1,DGH2,ZS} in order to relate \eqref{KdV5} to some known integrable equations by a Near-Identity Transformation (NIT) (also known as a Kodama transform \cite{KOD,KOD1}). We demonstrate that \eqref{KdV5} can be asymptotically matched to integrable equations of two separate types: (three) integrable equations of the same form as \eqref{KdV5}, with appropriate coefficients \cite{KK,KO,SK}; and the non-evolutionary Camassa-Holm \cite{CH} and Degasperis-Procesi \cite{DP} equations.

\subsection{ Near-identity transformation to integrable equations}
In this section we introduce the so-called {\it Near-Identity Transformation} (NIT) of the dependent variable $\eta(x,t).$  By employing this transformation, one can relate two HKdV equations of the type \eqref{KdV5} with different coefficients in terms of the order $\varepsilon^2$. This way, the model equation can be related to any one of the known integrable equations of the corresponding type.
Recall that $\eta(x,t)$ satisfies equation \eqref{KdV5} with known constant coefficients 
$c,$ $A,$ $\tilde{A},$ $M,$ $Q,$ $N_1,$ $N_2,$ given by \eqref{A}--\eqref{Q}. 
The NIT relates the $\eta$ function of  \eqref{KdV5} to another function $E(x,t)$  by way of
\begin{equation} \label{NIT}
    \eta(x,t)=E+\varepsilon( a_1E^2 + a_2 E_{xx}+a_3 E_x \partial_x^{-1}E),
\end{equation}
where the $a_i$ are 3 constant parameters and inverse differentiation means integration. This transformation is also known as the Kodama transform \cite{KOD1,KOD}, and appears in previous studies like \cite{DGH1,DGH,ZS}.
The relation between the two equations could be established as follows. Differentiating \eqref{NIT} we obtain
\begin{equation}
    \eta_t + c \eta_x  = E_t+ c E_x  + \mathcal{O}(\varepsilon)
\end{equation}
and since it is evident from \eqref{KdV5} that $ \eta_t + c \eta_x  =  \mathcal{O}(\varepsilon),$ then
\begin{equation} \label{E0}
     E_t+ c E_x =  \mathcal{O}(\varepsilon).
\end{equation}
Similarly, from  \eqref{NIT} and \eqref{E0} we have 
\begin{equation}
    \eta_t + c \eta_x + \varepsilon A \eta \eta_x + \varepsilon \tilde{A} \eta_{xxx}  = E_t+ c E_x +  \varepsilon A E E_x  + \varepsilon \tilde{A} E_{xxx}+ \mathcal{O}(\varepsilon^2).
\end{equation}
Equation  \eqref{KdV5} implies that $ \eta_t + c \eta_x  + \varepsilon A \eta \eta_x + \varepsilon \tilde{A} \eta_{xxx} =  \mathcal{O}(\varepsilon^2),$ therefore
\begin{equation}\label{xxx}
     E_t+ c E_x + \varepsilon A E E_x  + \varepsilon \tilde{A} E_{xxx}=  \mathcal{O}(\varepsilon^2).
\end{equation}
Thus the NIT does not change the coefficients of the terms up to the order $\varepsilon.$
Applying the \eqref{NIT} to the full equation \eqref{KdV5}, after some straightforward calculations involving \eqref{xxx}, one can verify that (up to terms of order $\varepsilon^2$) the associated evolution equation for $E$ is 
\begin{align}
    E_t+&c E_x+\varepsilon A E E_x + \varepsilon \tilde{A} E_{xxx} \notag \\
    &+\varepsilon^2 M' E^2 E_x +\varepsilon^2 Q' E_{5x}+\varepsilon^2(N'_1 E_x E_{xx}+N'_2 E E_{xxx})= \mathcal{O}(\varepsilon^3),
    \label{GKdV5E}
\end{align} 
where the $\mathcal{O}(\varepsilon^2) $ terms have the following coefficients (all terms involving $\partial_x ^{-1}E$  miraculously cancelling out):
\begin{equation} \label{MMprim}
\begin{split}
    M'&=M+A\left(a_1+\frac{1}{2}a_3\right), \\
    Q'&=Q, \\
    N_1'&=N_1+6\tilde{A} a_1-2A a_2+3 \tilde{A} a_3, \\
    N_2'&=N_2+3  \tilde{A} a_3. 
    \end{split}
\end{equation}

\subsection{Integrable KdV5-type equations}
The following equation of KdV5 type: 
\begin{equation}\label{KdVR}
    E_t + E_{ xxxxx } + 2 ( 6b+1 )E_{ x } E_{ xx } + 4 (b+1) E E_{ xxx } + 20b E^2 E_{ x } =0
\end{equation}
is known to be integrable for certain values of $b$. For instance,  when $b=4$ this equation corresponds to the Kaup-Kuperschmidt equation, \cite{KK}. Another integrable case is $b=\frac{1}{4}$, for which the equation is known as the Sawada-Kotera equation, \cite{SK} or Caudrey-Dodd-Gibbon equation \cite{CDG}, see also one of the early works \cite{SKa}. Finally, when $b=\frac{3}{2}$ the equation is the second equation from the integrable KdV hierarchy.  These three cases represent the only possible choices of coefficients leading to integrable equations in the given form -- see for example the classification results in \cite{DriSok,Mikh}.  Integrable equations have the advantage of possessing so-called soliton solutions, which are usually stable (in time) solitary waves that interact elastically, and recover their initial shape after interaction.  Soliton solutions for integrable equation can be obtained explicitly by various methods, such as the inverse scattering method \cite{AKNS,KK,SKa,ZMNP}. 

Our aim is to appropriately rescale the space and time variables in \eqref{KdVR}, as well as the the dependent variable $E$, so that we can relate \eqref{KdVR} to \eqref{KdV5} by employing the NIT.  
First, we change variables as follows:
\[
x\to x + \mathcal{C}_1 t,\quad t\to t\quad E\to E+\mathcal{C}_2
\] for some constants $\mathcal{C}_1,\mathcal{C}_2$. In terms of the new variables, we have:
\begin{multline}
   E_t + \left( 20b \mathcal{C}_2^2   + \mathcal{C}_1 \right) E_x  + E_{ xxxxx } + 2 ( 6b+1 )E_{ x } E_{ xx } + 4 (b+1) ( E+ \mathcal{C}_2 ) E_{ xxx }\\ + 20b E^2 E_{ x } + 40b \mathcal{C}_2 E E_{ x }   =0.
\end{multline}
   We further rescale the variables, introducing a scale parameter $\varepsilon$ as follows
\[
x\to  \frac{x}{\sqrt{\eps} \vartheta}, \quad t\to  \frac{t}{\sqrt{\eps} \vartheta}, \quad E\to \eps  \mathcal{ K } E
\]
and we obtain  
\begin{multline}\label{KdVaux}
   E_t + \left( 20b \mathcal{C}_2^2    + \mathcal{C}_1 \right) E_x  + \eps \left\{ 4 \vartheta^2 (b+1)    \mathcal{C}_2  E_{ xxx } + 40b   \mathcal{C}_2  \mathcal{ K } E E_{ x } \right\} \\  + \eps^2 \left\{ 4  \vartheta^2 (b+1)  \mathcal{K}   E E_{ xxx } + 2  \vartheta^2 ( 6b+1 )   \mathcal{ K } E_{ x } E_{ xx }  + 20b   \mathcal{ K }^2  E^2 E_{ x }  +  \vartheta^4  E_{ xxxxx } \right\}   = 0 .
\end{multline}
 This is an equation of the same type as \eqref{KdV5} with coefficients 
\begin{equation}\label{KdVK1}
\begin{split}
c &=  20b \mathcal{C}_2^2    + \mathcal{C}_1, \\
\tilde{A} &=  4 \vartheta^2 (b+1)    \mathcal{C}_2,  \\
A& =  40b   \mathcal{C}_2  \mathcal{ K },  \\
M'& = 20b   \mathcal{ K }^2  \\
N_1' &=  2  \vartheta^2 ( 6b+1 )   \mathcal{ K }  \\
N'_2 &= 4  \vartheta^2 (b+1)  \mathcal{K}  \\
Q'&=\vartheta^4 .
\end{split}
\end{equation}
From \eqref{KdVK1}, and the relation \eqref{MMprim} between the unknown $M', N_1', N_2', Q'$ and the known quantities $M,  N_1, N_2, Q$ (given by \eqref{A}--\eqref{Q}), we obtain
\begin{equation} \label{Mprim}
\begin{split}
       20b   \mathcal{ K }^2 &= M  + A   \left( a_1 + \frac{1}{2} a_3 \right),  \\
     4  \vartheta^2 (b+1)  \mathcal{K}  & =  N_2  + 3 \tilde A   a_3,  \\
     2  \vartheta^2 ( 6b+1 )   \mathcal{ K } &=  N_1 + 6 \tilde A  a_1 - 2 A  a_2 + 3 \tilde A   a_3,  \\
    \vartheta^4 &=Q.
    \end{split}
\end{equation} 
Then, from \eqref{KdVK1} and \eqref{Mprim} we find 
\[
 \vartheta = Q^{ \frac{1}{4} }, \quad \mathcal{C}_2 = \frac { \tilde A } { 4 (b+1) \sqrt{ Q }}, \quad \mathcal{K} = \frac{  (b+1) \sqrt{ Q }  A }{ 10 b  \tilde A  } 
\] and also
\[
\mathcal{C}_1 = c-\frac{5b \tilde{A}^2}{4(b+1)^2 Q}.
\]
It remains to determine the coefficients $a_1,a_2,a_3$ in the NIT. From \eqref{Mprim} we obtain
\begin{align}
    \label{aux1}
 a_1 &= \frac {  4   (b+1)^2  A^2 Q + 5 b A \tilde A N_2 - 30 b M \tilde A^2   } { 30 b A \tilde A^2 },   \\
 a_2 &= \frac {     (b+1)  A^2 Q +  b A \tilde A N_1 - 6 b M \tilde A^2   } { 2 b A^2 \tilde A },   \\
 a_3 &= \frac {  2   (b+1)^2  A Q   } { 15 b \tilde A^2 }-\frac {   N_2 } { 3 \tilde A }.   
\end{align}
The coefficients of the integrable \eqref{KdVaux} in terms of the known quantities are 
\begin{align}\label{aux2}
 M'  &= \frac {    (b+1)^2  A^2 Q   } { 5 b \tilde A^2 },   \\
  N'_1 &= \frac {   (b+1)  (6b+1)  A Q } { 5 b  \tilde A },   \\
 N'_2 &= \frac {  2   (b+1)^2  A Q   } { 5 b \tilde A }, \\
Q'&=Q.
\end{align}
In the formal limit $\gamma, \gamma_1,\rho_1,\omega \to 0$, which corresponds to a single layer irrotational fluid domain, the coefficients of \eqref{KdV5} are see also \cite{IV24}):
\begin{align}
     A  &= \frac{3 c}{2 h}, \,\,  \tilde A =\frac{c h^2}{6}, \,\,  Q  = \frac{19 c h^4}{360}, \,\,
 M  =-\frac{3 c}{8 h^2}, \,\, N_1  = \frac{23 c h}{24}, \,\,  N_2  =\frac{5 c h}{12}.
\end{align}
In this limit, \eqref{aux1} and \eqref{aux2} imply the coefficients of the integrable equation \eqref{KdVK1} are:
\begin{align}
      Q' &=  Q =\frac{19 c h^4}{360}, \,\,  M'=\frac{171 (b + 1)^2 c}{200 b h^2}, \, \, N_1' =\frac{19(6b + 1)(b + 1)ch}{200  b} ,\,\,  N_2'=\frac{19(b + 1)^2 ch}{100 b}.
\end{align}
Finally, the corresponding limits of the parameters in \eqref{aux1} are
\begin{align}
    a_1&= \frac{57 b^2 + 214 b + 57}{150 b h}, \,\, a_2=\frac{(202b + 57)h^2}{360 b }, \,\, 
    a_3=\frac{57 b^2 - 11b + 57}{150  b h }.
\end{align}
The soliton solutions of the integrable equations from this subsection could be written in terms of the so-called $\tau$-function as $E(x,t)=( \ln \tau(x,t))_{xx}, $ see \cite{CDG,SK}.
Thus $\partial^{-1}E=( \ln \tau(x,t))_{x}$ and therefore the NIT expression \eqref{NIT} is easier to find in terms of $\tau(x,t),$ which can be expressed in terms of the soliton parameters (scattering data).

\subsection{Camassa-Holm and Degasperis-Procesi equations}
The Camassa-Holm (CH) equation was found after its derivation as a shallow water equation in \cite{CH} to fit into a class of integrable equations derived previously by using hereditary symmetries in Fokas and Fuchssteiner \cite{FF}.
The Degasperis–Procesi (DP) equation \cite{DP} was discovered in a search for integrable equations similar in form to the Camassa–Holm equation, the Lax representation was discovered in \cite{DHH}. Both equations have some remarkable properties which attracted an enormous interest towards them. Both equations can be written in the following form (cf. \cite{IV}) 
\begin{align}\label{eq01}
  \tilde{U}_T= & \Gamma_1 \tilde{U}_X+ \vartheta^2  \tilde{U}_{XXT}+\Gamma_2 \tilde{U}_{XXX}-  \vartheta \mathcal{ B } \Upsilon (b+1)  \tilde{U}  \tilde{U}_X +  \vartheta^3 \mathcal{ B } \Upsilon  (b \tilde{U}_X  \tilde{U}_{XX}+ \tilde{U} \tilde{U}_{XXX})
\end{align} where the CH equation corresponds to the parameter choice $b=2$ and the DP equation corresponds to $b=3,$  and $\vartheta, \mathcal{ B }, \Upsilon, \Gamma_1,\Gamma_2$ are arbitrary constants. There are no other values of $b$ for which the equation \eqref{eq01} is integrable  \cite{Iv05}.
Note that the quantities in this section are not related to the quantities bearing the same notation in the previous subsection.  
The scale parameter $\varepsilon$ may be introduced via
\begin{equation}
   X= \frac{x}{\sqrt{\varepsilon}}, \quad T= \frac{t}{\sqrt{\varepsilon}}, \quad \tilde{U}=\varepsilon U,
\end{equation}
\begin{align}\label{eq01a}
  U_t= & \Gamma_1 U_x+ \varepsilon (\vartheta^2 U_{xxt}+\Gamma_2U_{xxx})- \varepsilon \vartheta \mathcal{ B } \Upsilon (b+1) UU_x + \varepsilon^2 \vartheta^3 \mathcal{ B } \Upsilon  (bU_xU_{xx}+UU_{xxx}).
\end{align}
In what follows our approach differs from that of \cite{DGH1,DGH2,DGH}, which uses the Helmholtz operator and from the derivation in \cite{JO}. In  expression \eqref{eq01a} we substitute $\varepsilon \vartheta^2 U_{xxt}=\varepsilon \vartheta^2[ U_t]_{xx}$, with $U_t$ given by \eqref{eq01a}:
\begin{align}\label{eq02}
  \varepsilon \vartheta^2 U_{xxt}&=\varepsilon \vartheta^2[\Gamma_1 U_x+ \varepsilon (\vartheta^2 U_{xxt}+\Gamma_2U_{xxx})- \varepsilon \vartheta \mathcal{B} \Upsilon (b+1) UU_x]_{xx} +...\nonumber \\
  &= \varepsilon \vartheta^2\Gamma_1 U_{xxx} + \varepsilon^2 (\vartheta^4 U_{xxxxt}+\vartheta^2 \Gamma_2U_{xxxxx}) - \varepsilon^2 \vartheta ^3 \mathcal{B} \Upsilon (b+1) [UU_x]_{xx} +... \nonumber \\
  &=\varepsilon \vartheta^2\Gamma_1 U_{xxx} + \varepsilon^2 [\vartheta^4 (\Gamma_1U_x)_{xxxx}+\vartheta^2 \Gamma_2U_{xxxxx}] - \varepsilon^2 \vartheta ^3 \mathcal{B} \Upsilon (b+1) [UU_x]_{xx} +... \nonumber \\
   &=\varepsilon \vartheta^2\Gamma_1 U_{xxx} + \varepsilon^2 (\vartheta^4 \Gamma_1+\vartheta^2 \Gamma_2)U_{5x } - \varepsilon^2 \vartheta ^3 \mathcal{B} \Upsilon (b+1) [3U_xU_{xx}+UU_{xxx}] +..., \quad
\end{align}
with neglected terms of order $\varepsilon^3$.
Plugging the asymptotic expression for $ \varepsilon \vartheta^2 U_{xxt}$ from \eqref{eq02} back into \eqref{eq01a} gives
\begin{align}\label{eq03}
  U_t= & \Gamma_1 U_x+ \varepsilon ( \vartheta^2\Gamma_1 +\Gamma_2)U_{xxx} \nonumber \\
  & + \varepsilon^2 (\vartheta^4 \Gamma_1+\vartheta^2 \Gamma_2)U_{5x } - \varepsilon^2 \vartheta ^3 \mathcal{B} \Upsilon (b+1) [3U_xU_{xx}+UU_{xxx}] \nonumber \\
  &- \varepsilon \vartheta \mathcal{B} \Upsilon (b+1) UU_x + \varepsilon^2 \vartheta^3 \mathcal{B} \Upsilon (bU_xU_{xx}+UU_{xxx})+\mathcal{O}(\varepsilon^3)
\end{align}
or
\begin{align}\label{eq05a}
  U_t- & \Gamma_1 U_x- \varepsilon ( \vartheta^2\Gamma_1 +\Gamma_2)U_{xxx} + \varepsilon \vartheta \mathcal{B} \Upsilon (b+1) UU_x   - \varepsilon^2 \vartheta^2(\vartheta^2 \Gamma_1+\Gamma_2)U_{5x }            \nonumber \\
  & + \varepsilon^2 \varepsilon ^3 \mathcal{B} \Upsilon  [(2b+3)U_xU_{xx}+bUU_{xxx}]=\mathcal{O}(\varepsilon^3).
\end{align}
This now needs to be related to the physical equation \eqref{KdV5}, 
with known coefficients $c, A,\tilde{A},Q,M, N_1,N_2.$
We take $\Upsilon=1$ so that $\eta$ is expressed as a NIT  of $U$, 
\begin{equation}\label{NIT2}
    \eta = U+\varepsilon(a_1 U^2+a_2 U_{xx}+a_3 U_x \partial^{-1} U) .
    \end{equation}
Note that $U$ is a solution of an integrable equation (CH or DP, as in \eqref{eq01a}), while $\eta$ satisfies the non-integrable physical model equation \eqref{KdV5}. However, the integrable model \eqref{eq01a} is asymptotically equivalent to a HKdV model with coefficients as in \eqref{eq05a}. 
Thus, we identify $c=-\Gamma_1 $ and $\tilde{A}=  -( \vartheta^2\Gamma_1 +\Gamma_2),$ $ A=\vartheta \mathcal{B} \Upsilon (b+1),$ $$Q'=Q=-\vartheta^2(\vartheta^2 \Gamma_1+\Gamma_2),$$
Therefore, $\Gamma_1=-c,$ and 
$$ \vartheta^2 =\frac{Q}{\tilde{A}}, $$ thus
$$\Gamma_2 = \vartheta^2 c - \tilde{A} = \frac{Qc-\tilde{A}^2}{\tilde{A}}.$$
Also $\vartheta \mathcal{B} \Upsilon (b+1)=A$
and since $\Upsilon=1$ we have $$\mathcal{B} = \frac{A}{\vartheta(b+1)}=\frac{A\sqrt{\tilde{A}}}{\sqrt{Q}(b+1)}.$$ 
 We need to  also determine the $a_i$, which are the yet unknown coefficients of the NIT.
 The remaining relations are
 \begin{align}
 M'&= 0=M+A\left(a_1+\frac{1}{2}a_3\right)=0  \label{Meq} \\
  \nonumber N_1' &= \vartheta^3 \mathcal{B} \Upsilon (2b+3)= N_1+6\tilde{A} a_1-2Aa_2+3\tilde{A} a_3, \\
\nonumber   N_2' &=\vartheta^3 \mathcal{B} \Upsilon b = N_2+3\tilde{A} a_3,
 \end{align}
and  from the last equation we get $$a_3=\frac{1}{3\tilde{A} }\left(\frac{AQb}{\tilde{A} (b+1)}-N_2\right),$$
 We obtain $a_1$ from \eqref{Meq}, and  $a_2$ from the remaining relation, to get
\[
     a_1= \frac{N_2}{6 \tilde{A}}-\frac{AQb}{6\tilde{A}^2 (b+1)}-\frac{M}{A}, \qquad
     a_2 = \frac{N_1}{2A} - \frac{3 \tilde{A}M}{A^2}-\frac{(2b+3) Q}{2(b+1)\tilde{A}}.
\]
The integrable equations \eqref{eq01} become ($b=2,3$)
\begin{align}\label{eq01f}
  (\tilde{U}_T -\frac{Q}{\tilde{A}}\tilde{U}_{XX})_T+ c\tilde{U}_X-\frac{Qc-\tilde{A}^2}{\tilde{A}} \tilde{U}_{XXX}+  A  \tilde{U}  \tilde{U}_X - \frac{AQ}{(b+1)\tilde{A}} (b \tilde{U}_X  \tilde{U}_{XX}+ \tilde{U} \tilde{U}_{XXX})=0.
\end{align}
Taking the limiting cases $\gamma, \gamma_1,\rho_1,\omega \to 0$, which corresponds physically to one layer of irrotational fluid, leads to the equation
\begin{align}\label{eq01f1}
  (\tilde{U}_T -\frac{19 h^2}{60}\tilde{U}_{XX})_T+ c\tilde{U}_X-\frac{ 3 c h^2 }{20}\tilde{U}_{XXX}+  \frac{3 c}{2 h}  \tilde{U}  \tilde{U}_X - \frac{19 c h}{40 (b + 1)} (b \tilde{U}_X  \tilde{U}_{XX}+ \tilde{U} \tilde{U}_{XXX})=0,
\end{align} with NIT coefficients
\begin{align}
    a_1 &=\frac{23 b + 80}{120 h  (b + 1)}, \quad    a_2= \frac{h^2 (31 b - 26)}{360 (b +1)}, \quad    a_3 = \frac{7b - 50}{60 h (b + 1)}.
\end{align}
The CH and DP soliton solutions can be written in parametric form  \cite{CoIv,ILO,Mats1,Mats2,CGI} as
\begin{equation}
    U(X(y,t),t)=X_t(y,t), \quad x=X(y,t),
\end{equation}
for $y$ a parameter, and $X$ given in terms of the soliton parameters (scattering data). It should be mentioned that the smooth soliton solutions do not always exist for any smooth initial data. For some types of smooth initial data the Inverse Scattering Method needs substantial modifications as shown for example in \cite{K}.

\section{Conclusions and Discussion}
 In this paper we have demonstrated how the nonlinear governing equations for internal wave-current interactions can be re-expressed as a closed system of coupled equations given in terms of Dirichlet-Neumann (DN) operators for the upper and lower fluid layers. Following this, we derived a Higher-order KdV  model equation for the system using small-amplitude,  and long-wave, asymptotic approximations in a regime  which satisfies the KdV-type parameter balance: $\epsilon\sim\delta^2$.
 We have demonstrated how the Higher order KdV model equation \eqref{KdV5} can be related to several celebrated integrable model equations by way of the Near-Identity Transform, the solutions of which are well studied. The integrable KdV-5 equation is from the KdV hierarchy, and accordingly its solitons can be recovered from those of the KdV equation. Furthermore, inverse scattering theory and the soliton solutions of the Kaup-Kuperschmidt equation have been studied in \cite{KK,VSG}, and similarly for the Sawada-Kotera equation in \cite{SK,KK,CDG,SKa}.
Both the CH and DP equations possess famed `peakon' solutions, however the existence of such solutions would require $\Gamma_1=\Gamma_2=0$ in our physical model. The existence of solutions corresponding to the CH and DP `cuspons' is also not a straightforward matter for this HKdV model equation, since the NIT \eqref{NIT2} involves $x$-derivatives of $U$ and would therefore become singular.

We have a special case when $A=0$. Then the HKdV model as presented here requires adjustment, since the leading order nonlinearity is $\eta^2 \eta_x,$ which is balanced by the $\eta_{xxx}$ dispersive term. In this case  the appropriate scaling requires $\varepsilon$ and $\delta$ to be of the same order, and the corresponding model equation is the Gardner equation, see  \cite{CuIv,IMT}.

 One of the limitations of the model is the flat surface approximation, which treats the internal waves as shallow water waves and neglects the surface waves. However, in addition to this approximation, the internal waves on the thermocline can be considered as playing the role of bathymetry for the (much smaller and faster) surface waves, which could be treated again as classical water waves by using the appropriate Dirichlet-Neumann operator (DNO) for the upper fluid domain. Moreover, in the framework of the so-called thermal shallow-water model (TSWM, see for example \cite{Del,Zet}) one may relax the homogeneity assumption of the standard shallow water theory, and allow the fluid properties within each layer to vary in the horizontal direction. This approach considers an {\it active} upper layer, which is characterised by a spatially varying density, that is
$$\rho_1(x,t)=\rho+ \Delta \rho (x,t),$$ for a bounded density contrast $\Delta \rho(x,t)$ and a much
deeper, quiescent lower layer with constant density $\rho.$ The equivalent barotropic
approximation \cite{Gill} then leads to a single set of shallow water equations (the TSWM) for the active upper layer.

\subsection*{Acknowledgements} The authors are thankful to the anonymous referees for their helpful comments and suggestions. This publication has resulted from research conducted with funding from the Science Foundation Ireland under Grant number 21/FFP-A/9150.

\subsection*{Conflict of Interest} None of the authors have a conflict of interest to disclose.

\section{Appendix} 
\subsection{Coefficients of the lower order  terms}
First, matching the terms containing $\eta_x$ in \eqref{compat}, by taking into account equations \eqref{AUX1a} and \eqref{AUX2b} and recalling \eqref{wavespeed}, we have indeed $ A_1 =  \frac c { \alpha_1}$. 
In the same manner, matching the terms containing $\eta \eta_x$ we have
\begin{align*}
B_1 =  \frac { -  \alpha_3  c + \alpha_1 c (   \alpha_4    - 2  \Gamma    \alpha_3  ) +  \alpha_1^2  ( \alpha_6 -  \Gamma  \alpha_4  )}{ 2  \alpha_1^2 ( 2 c +  \Gamma  \alpha_1 )}=   -  \frac {  \alpha_3  c  }{   \alpha_1^2 } -   \frac {   \alpha_4    }{ 2  \alpha_1 } + \frac { 3 \alpha_3  c^2 + 3 \alpha_1  \alpha_4 c + \alpha_6 \alpha_1^2 } { 2 \alpha_1^2 ( 2 c + \Gamma \alpha_1 ) }.
\end{align*}
The corresponding coefficient  in \eqref{KdV5} can be found from \eqref{AUX1a} and  \eqref{etaf1}, and with the above expression for $B_1$:
\begin{align}
A  = & 2 \alpha_1 B_1 +  2 \alpha_3 A_1 + \alpha_4  = \frac { 3 \alpha_3  c^2 + 3 \alpha_1  \alpha_4 c + \alpha_6 \alpha_1^2 } { \alpha_1 ( 2 c + \Gamma \alpha_1 ) }   \nonumber \\
& = \frac {  \alpha_1   } { h^2 h_1^2  ( 2 c + \Gamma \alpha_1 )   }\left\{   \rho  h_1^2  \left[    \frac {  3 } { 4 }  c^2  + (  \frac {   3 } { 2 } c + \gamma  h )^2 \right]   - \rho_1 h^2 \left[  \frac {  3 } { 4 }   c^2 + (  \frac {   3 } { 2 }  c - \gamma_1 h_1 )^2 \right]     \right\}.
\nonumber \end{align}
We proceed by examining the coefficients of the other terms of order $\eps$ in \eqref{ansatz} and \eqref{KdV5}. In particular, we now match the $\eta_{xxx}$ terms in order to obtain: 
 \[ 
B_2 =  \frac { -  \alpha_2  c   (  c + \Gamma \alpha_1  ) }{  \alpha_1^2 ( 2 c +  \Gamma  \alpha_1 )} = - \frac {   \alpha_2  c    }{  \alpha_1^2 } + \frac {   \alpha_2  c^2   }{  \alpha_1^2 ( 2 c +  \Gamma  \alpha_1 ) },
\]
and the corresponding coefficient from \eqref{AUX1a} is:
\begin{align}
&\tilde A  =  \alpha_1 B_2 +   \alpha_2 A_1 = \frac {  \alpha_2 c^2 } { \alpha_1 ( 2 c + \Gamma \alpha_1 ) } =  \frac 1 3   \frac {  (  \rho h +  \rho_1 h_1 )  c^2 } {  ( 2 c + \Gamma \alpha_1 ) } \alpha_1. \nonumber
\end{align}
At this level of approximation we recover the coefficients of a KdV equation,  as in \cite{IC,CIT}.

\subsection{Coefficients of the higher order terms}
We follow the same strategy for the higher-order terms, that is, we use \eqref{compat}, \eqref{AUX1a} and \eqref{AUX2b} to determine the constants in \eqref{ansatz}, with the coefficients in \eqref{KdV5} following from \eqref{etaf1}.
First, for the coefficients of $\eta^3 $ in \eqref{ansatz}, we have
 \begin{align*}
 C_1  =  -   \frac { \alpha_3 B_1 } {  \alpha_1 } + \frac { \beta_1  c } { \alpha_1^2 } + \frac {  \beta_5 } { \alpha_1 }  + \frac { 2  \alpha_1 B_1  \left(  2 \alpha_3 c + \alpha_1 \alpha_4   - 4  \alpha_1^2 B_1 \right)   -   12 \beta_1   c^2  - 24 \alpha_1 \beta_5 c   -  3 \alpha_1^2  \beta_7   } { 6 \alpha_1^2   ( 2 c + \Gamma \alpha_1 ) }.  
\end{align*}
The  latter allows us to express the coefficient of $\eta^2 \eta_x$ in \eqref{KdV5} as follows:
\begin{align}
 M = &3 ( C_1 \alpha_1 +  \alpha_3 B_1 -  \beta_1  A_1 -  \beta_5 )  \nonumber \\
&= 3\left [    \alpha_1   B_1 \frac { 2 \alpha_3 A_1 + \alpha_4   - 4  \alpha_1 B_1  } { 3 ( 2 c + \Gamma \alpha_1 ) }  -  \frac { 2 \beta_1  A_1 c } { 2 c + \Gamma \alpha_1}  - \frac { 4 \beta_5 c } { 2 c + \Gamma \alpha_1}  - \frac {  \beta_7 \alpha_1  } { 2 ( 2 c + \Gamma \alpha_1 ) }  \right].
\nonumber \end{align}
 For the coefficients of the term $\eta \eta_{xx}$, in \eqref{ansatz}, we have
  \begin{align*}
  C_2  = -  \frac{ 2 \alpha_2 B_1 } { \alpha_1 } -  \frac { \alpha_3 B_2 } { \alpha_1 } - \frac { \beta_2 c } {  \alpha_1^2  } +  \frac { 3 \beta_3 c  } { \alpha_1^2 } + \frac { 2 \beta_6 } { \alpha_1 }    + \frac { - 4 \alpha_1^3   B_1 B_2 +  \beta_2   c^2  - 4 \beta_3  c^2     -  4  \alpha_1 \beta_6 c  } { \alpha_1^2 ( 2 c + \Gamma \alpha_1 ) }.
\end{align*}
Then, the corresponding coefficient  in \eqref{KdV5} is given by:
\begin{align}
 N_2 = &\alpha_1 C_2 + 2 \alpha_2 B_1 + \alpha_3 B_2 + \beta_2 A_1 - 3 \beta_3 A_1 - 2 \beta_6  \nonumber \\
&= -   \frac { 4 \alpha_1^2 B_1 B_2 } {  2 c + \Gamma \alpha_1  } +  \frac {  \beta_2  A_1 c } { 2 c + \Gamma \alpha_1}  -  \frac { 4 \beta_3  A_1 c } { 2 c + \Gamma \alpha_1}  - \frac { 4 \beta_6 c } { 2 c + \Gamma \alpha_1}.
\nonumber \end{align}
A lengthy calculation shows that the coefficient of $\eta_{x}^2$ in \eqref{ansatz} is given by:
\begin{align*}
&  C_3  = - \frac {   C_2 }  { 2  } - \frac { 3 \alpha_2  B_1 } {  \alpha_1 }  - \frac {   \alpha_3 B_2  } { 2 \alpha_1 } - \frac  { 3 \beta_2 c  } { 2 \alpha_1^2 }  + \frac { 7 \beta_3 c } { 2 \alpha_1^2 }  +  \frac { 3 \beta_6 } {  \alpha_1 }   \\
& +    \frac {   \beta_2  A_1 c - 4  \beta_3  A_1 c  - 4  \beta_6 c + 3  \alpha_1  \alpha_2  A_1 B_1 - 3 \alpha_1^2   B_1 B_2 - 2 \alpha_1 \alpha_3   A_1 B_2 -   \alpha_1 \alpha_4    B_2 } {  \alpha_1 ( 2 c + \Gamma \alpha_1 ) } .
\end{align*}
Using the last expression, we have
\begin{align}
 N_1 =& \alpha_1 C_2 + 2 \alpha_1 C_3 + 6 \alpha_2  B_1 +  \alpha_3 B_2 +3 \beta_2 A_1 - 7 \beta_3 A_1 - 6 \beta_6   \nonumber \\
& = \frac {2   \beta_2   c^2   -   8  \beta_3  c^2   -   8 \alpha_1  \beta_6 c      +  8 \alpha_1   \alpha_2  B_1 c   - 2 \alpha_1^2 \left(    B_1 \tilde A  +  B_2 A   \right ) }   { \alpha_1 ( 2 c + \Gamma \alpha_1 ) }. \nonumber
\end{align}

Finally, for the coefficients of $\eta_{xxxx}$ in \eqref{ansatz}, we have
\begin{align*}
C_4 & = - \frac { B_2 \left[ B_2 \alpha_1^3 + (  2 c + \Gamma \alpha_1 ) \alpha_1 \alpha_2 \right] +  \beta_4 c ( c + \Gamma \alpha_1 )} { \alpha_1^2 ( 2 c + \Gamma \alpha_1)}   = - \frac { \alpha_2 B_2 } { \alpha_1 } - \frac{  \beta_4 c } { \alpha_1^2 } +  \frac {  \beta_4  c^2  -  \alpha_1^3 B_2^2 } {  \alpha_1^2 ( 2 c + \Gamma \alpha_1 )  }.
\end{align*}
We can then determine the remaining term in \eqref{KdV5}:
 \begin{align}
  Q=& \alpha_1  C_4  + \alpha_2 B_2 + \beta_4 A_1    = \frac {  \beta_4  c^2  -  \alpha_1^3 B_2^2 } {  \alpha_1 ( 2 c + \Gamma \alpha_1 )  }. \nonumber \end{align}
Thus, we have determined all coefficients in \eqref{KdV5} in terms of the given constants. These expressions agree with the coefficients appearing in \cite{RW,GKK}, which have been derived in a different physical context.

\end{document}